\documentclass[showpacs,showkeys,preprint,amsmath,amssymb,nofootinbib, prd, aps, 11pt]{revtex4-1}
\usepackage[T1]{fontenc}
\usepackage[latin1]{inputenc}

\usepackage[sc]{mathpazo}
\usepackage{framed, color}
\definecolor{shadecolor}{rgb}{0.84,0.84,0.84}
\usepackage{graphicx}
\newcommand{\Det}{\mbox{Det}}
\newcommand{\tr}{\mbox{tr}}

\newcommand{\vek}[1]{\mathbf{#1}}

\newcommand{\wb}{\bar{\omega}_{\perp}}
\newcommand{\tb}{\bar{\tau}}

\newcommand{\WW}{\mathcal{W}}
\newcommand{\Wb}{\overline{\WW}}

\newcommand{\JJ}{\mathcal{J}}
\newcommand{\NN}{\mathcal{N}}

\newcommand{\ka}{k}

\newcommand{\qa}{q}
\newcommand{\ks}{k}

\newcommand{\qs}{q}

\newcommand{\PT}{\mathcal{P}^{\perp}}
\newcommand{\PL}{\mathcal{P}^{\|}}

\newcommand{\kargs}{k_0,\vek{k}}

\newcommand{\unitthree}[1]{\widehat{\mathbf{#1}}}
\newcommand{\unitfour}[1]{\widehat{#1}}
\newcommand{\dd}{\text{\dj}}
\newcommand{\wt}{\omega_\perp}
\newcommand{\wl}{\omega_\|}
\newcommand{\xt}{\chi_\perp}
\newcommand{\xl}{\chi_\|}
\newcommand{\Mt}{M_{\perp}}
\newcommand{\Ml}{\Delta M_{\|}}

\renewcommand{\sb}{\bar{\omega}_{\|}}

\begin{document}

%%%%%%%%%%%%%%%%%%%%%%%%%%%%%%%%%%%%%%%%%%%%%%%%%%%%%%%%%%%%%%%%%%%%%%%%%%%%%%%%

% Force line breaks with \\
\title{A covariant variational approach to Yang-Mills Theory at finite temperatures}

\author{M.~Quandt}\email{markus.quandt@uni-tuebingen.de}
\author{H.~Reinhardt}\email{hugo.reinhardt@uni-tuebingen.de}
%\author{J.~Heffner}\email{jan.heffner@uni-tuebingen.de}
\affiliation{%
Universit\"at T\"ubingen\\
Institut f\"ur Theoretische Physik\\
Auf der Morgenstelle 14\\
D-72076 T\"ubingen, Germany
}%

\date{\today}% It is always \today, today,

%%%%%%%%%%%%%%%%%%%%%%%%%%%%%%%%%%%%%%%%%%%%%%%%%%%%%%%%%%%%%%%%%%%%%%%%%%%%%%%%

\begin{abstract}
We extend the covariant varitional approach for $SU(N)$ Yang-Mills theory 
in Landau gauge to non-zero temperatures. The renormalization of the 
zero-temperature case is revisited and it is shown that the same 
counterterms are sufficient to render the low-order Green's function finite
at non-zero temperature. We compute the ghost and gluon propagator 
numerically and show that it agrees in all qualitative respects 
with the results of high-precision lattice calculations.
\end{abstract}

%%%%%%%%%%%%%%%%%%%%%%%%%%%%%%%%%%%%%%%%%%%%%%%%%%%%%%%%%%%%%%%%%%%%%%%%%%%%%%

\pacs{11.80.Fv, 11.15.-q}% PACS, the Physics and Astronomy Classification Scheme.
\keywords{gauge theories, confinement, variational methods, Landau gauge} 
\maketitle

%%%%%%%%%%%%%%%%%%%%%%%%%%%%%%%%%%%%%%%%%%%%%%%%%%%%%%%%%%%%%%%%%%%%%%%%%%%%%%

\section{Introduction}
\label{intro}

\noindent
In recent years functional continuum methods have been extensively used to study 
the low energy sector and the phase diagram of quantum chromodynamics (QCD).
These methods include functional renormalization group (FRG) flow equations 
\cite{Fischer:2006ub,Alkofer:2000wg,Binosi:2009qm}, 
Dyson-Schwinger equations (DSE) \cite{Pawlowski:2005xe,Gies:2006wv} 
and variational methods \cite{Quandt:2013wna,Feuchter:2004mk,Feuchter:2004gb}. 
% Most analytical approaches to Yang-Mills theory concentrate on the low-order 
% Green's functions, since these objects are most directly accessible in such 
% a setup. 
The gauge-variance of the Green's functions makes it necessary to fix a gauge, 
and most techniques such as FRG and DSE initially concentrated their studies on 
the case of covariant gauges. 
This choice has a two-fold advantage: On the one hand, the BRST symmetry and 
the ensuing Slavnov-Taylor identities provide constraints to guide the 
analysis. More importantly, however, the Kuga-Ojima criterion
\cite{Ojima:1978hy,Kugo:1979gm} claims a direct connection, based on the 
BRST mechanism, between the propagators in Landau gauge and physical 
phenomena such as colour confinement. 

At the quantum level, it is not immediately clear if BRST symmetry is naively 
maintained or visible. Most of the functional studies quoted above initially found an 
infrared vanishing, \emph{scaling} type of solution for the gluon propagator 
(as the Kugo-Ojima criterion would suggest), which is, however, at odds with 
high-precision lattice simulations \cite{Cucchieri:2007md, Cucchieri:2007rg, 
Bogolubsky:2009dc,Sternbeck:2013zja}. It was later shown that infrared 
finite \emph{decoupling} solutions could also be obtained, if the infrared behaviour 
is sufficiently constrained \cite{Fischer:2008uz}; such solutions had also been 
found earlier \cite{Aguilar:2004sw, Aguilar:2007nf,Aguilar:2007ie,Binosi:2007pi,
Aguilar:2008xm,Boucaud:2006if,Boucaud:2008ji,Boucaud:2008ky,Dudal:2005na,Dudal:2007cw,
Capri:2007ix,Dudal:2008sp,Oliveira:2007dy,Oliveira:2008uf,RodriguezQuintero:2010wy,
Huber:2012kd,Pelaez:2014mxa}.
The decoupling solutions agree very well with lattice data, but indicate a (soft) BRST
breaking in the full theory.  

In Ref.~\cite{Quandt:2013wna}, we proposed a variational approach that is based on the 
effective action for the gluon propagator. The technique offers several conceptual advantages: 
For instance, it automatically yields a closed set of integral equations that can be 
renormalized through conventional counter terms without further truncation. In addition, 
the variational approach allows, in principle, to discern between the scaling and decoupling 
type of solution, as the one with the lower effective action must be realized. Numerically, 
the approach yields excellent agreement with lattice data that is on par with the best functional 
methods listed above. 

On the other hand, the variational approach violates BRST symmetry and the Kugo-Ojima criterion 
does not apply so that the question of colour confinment in the variational approach is mute at 
the moment. This explicit violation of BRST symmetry is unavoidable (to a certain extent) 
in \emph{any} analytical approach, and it is important that the breaking occurs in a controlled way.
For the variational approach, BRST symmetry is implemented exactly in the target functional and 
an unconstrained variation will give the exact solution. It is only through restrictions on the 
trial variation measure that the breaking of BRST symmetry occurs. As we enlarge the 
variational ansatz space, we  improve the quality of the dynamics, and at the same time 
reduce the violation of BRST symmetry (even though we do not have a reliable measure to 
quantify this).\footnote{This is consistent since symmetries are invariances of the 
dynamics, i.e.~it is not necessary nor appropriate to enforce a symmetry of the exact model 
onto the truncated dynamics.} The situation is, in fact, analogous to the Hamiltonian 
formulation in Coulomb gauge 
\cite{Schutte:1985sd, Szczepaniak:2001rg, Feuchter:2004mk,Reinhardt:2004mm, Epple:2006hv}
where Gauss' law can be implemented exactly in the target Hamiltonian, but only the 
exact solution of the functional Schr\"odinger equation will obey Gauss' law exactly.

Numerically, the variational solution of Ref.~\cite{Quandt:2013wna} 
describes the propagators very well, particularly in the 
mid-momentum region which is most important phenomenologically, expecially for the deconfinement 
phase transition. It is therefore natural to extend this approach to finite temperatures and 
finite chemichal potentials, and eventually include dynamical fermions. 
In the present paper, we describe the first step in this program, viz.~the 
introduction of non-zero temperature. Naturally, this extension has also 
been done in the FRG \cite{Pawlowski:2014aha} and DSE \cite{Welzbacher:2014pea}
as well as in a perturbative approach \cite{Reinosa:2014ooa,Reinosa:2014zta}. 
The lattice data \cite{Aouane:2011fv} show a cross-over type of 
signal in the ratio of the two inequivalent Lorentz structures for the gluon propagator, 
but no clear qualitative change at the deconfinement phase transition. In the present paper, 
we will concentrate on the finite-temperature propagators and search for signals of 
a phase transition that can be derived from them alone. Eventually, the present work should 
be extended to measure the effective action for the Polyakov loop, which is the 
real order parameter for Yang-Mills theory at finite temperature. This study will be 
reported elsewhere. 

The paper is organised as follows: In the next section, we revisit the covariant variational
principle and extend it to finite temperatures through the imaginary time formalism. 
In section \ref{sec:3} we compute the effective action for the gluon propgagator at 
finite temperature and derive the unrenormalized gap equation. Section \ref{sec:4}
simplifies and improves the renormalization at zero temperature discussed in Ref.~\cite{Quandt:2013wna}. 
We demonstrate that the same counter terms are also sufficient to render the theory finite at 
any non-zero temperature, and present the fully renormalized finite-temperature corrections to 
the integral equations. In section \ref{sec:5}, we report our numerical treatment and present 
solutions for the ghost and gluon propagator. In particular, we study possible signals for a 
phase transition in the gluon and ghost propagator. In the last section, we conclude this study 
with a brief summary and an outlook on further extensions of the variational method.

%%%%%%%%%%%%%%%%%%%%%%%%%%%%%%%%%%%%%%%%%%%%%%%%%%%%%%%%%%%%%%%%%%%%%%%%%%%%%%%%%%%%%%%%%%%%%%%%%%%%%%

\section{The variational principle}
\label{sec:2}

\subsection{The variational method at finite temperatures}
\label{sec:2a}
In Ref. \cite{Quandt:2013wna}, we described the basics of the covariant variational 
principle in quantum field theory: for a theory with a quantum field $A$ defined by an 
action $S(A)$ in Euclidean space time, the variation is over all normalised path integral 
(probability) measures $d\mu(A)$ used to compute quantum averages according to 
$\langle \cdots \rangle_\mu \equiv \int d\mu(A)\cdots$.
%It is important that all trial measures be \emph{probability measures} normalised
%according to  $\langle 1 \rangle_\mu =  \int d\mu(A) = 1$.
If the measure is written in Radon-Nikodym form $d\mu(A) = dA\,\rho(A)$ 
with a suitable density $\rho$, the \emph{entropy}
\begin{equation}
\WW(\mu) \equiv - \langle \ln \rho \rangle_\mu = 
- \int dA\,\rho(A)\,\ln \rho(A)
\label{1.1}
\end{equation}
describes the accessible field space for quantum fluctuations. In the full 
theory, the entropy of the fluctuations is balanced against the classical 
Euclidean action such that the \emph{free action}
\begin{equation}
F(\mu) \equiv \langle S\rangle_\mu - \hslash \WW(\mu) 
\stackrel{!}{=} \mbox{min}
\label{1.2}
\end{equation}
is minimized. This is because the unique solution of the variational principle 
(\ref{1.2}) is the Gibbs-like measure
\begin{align}
d\mu_0(A) = Z^{-1}\,\exp\left[ - \hslash^{-1}\,S(A)\right]\,dA 
\end{align}
whose moments are the conventional Schwinger functions; the minimal free action 
is simply $F( \mu_0) = -\ln Z$. The \emph{quantum effective action} 
is a constrained version of the free action,
\begin{equation}
\Gamma(\omega) \equiv  \min_\mu F(\mu,\omega) \equiv \min_\mu 
\Big\{ \langle S\rangle_\mu - \hslash \WW(\mu)\, \,\big|\,\,
\langle \Omega \rangle_\mu = \omega  \Big\}\,,
\label{1.4}
\end{equation}
where an arbitrary operator $\Omega$ is fixed to a prescribed external
value $\omega$. Usually, $\Omega$ is chosen as the quantum field operator $A$ 
itself whence $\omega$ becomes the \emph{classical field}, but this is not 
mandatory: for the present study, it is more convenient to take 
$\Omega$ as the $2$-point func\-tion of the quantum field and compute the optimal 
propagator from the overall minimisation of the effective action, 
$\delta \Gamma / \delta \omega = 0$.  

To transcribe this method to non-zero temperatures, we can follow the standard
imaginary time formalism and restrict the Euclidean time interval to $[0,\beta]$ with 
anti/periodic temporal boundary conditions for fermions/bosons, respectively.
This procedure works because the free action in our approach is always computed from a 
conventional quantum field theory, in which temperature can be introduced as described. 
The minimal free action eq.~(\ref{1.2}) obtained for the Gibbs measure is still 
$F(\mu_0) = -\ln Z$, but with the partition function 
\begin{align}
Z(\beta) &= \int_\beta dA\,\exp\left[ - \hslash^{-1}\,S(A)\right]
\end{align}
now computed using fields with the appropriate temporal boundary conditions. 
With the corresponding Fourier decomposition (in the bosonic case)
\begin{align}
A(t,\vek{x}) = \beta^{-1}\sum_{n \in \mathbb{Z}}\int \frac{d^3 k}{(2\pi)^3}\,
e^{i \,(\nu_n t + \vek{k}\vek{x})}\,A_n(\vek{k})\,,
\qquad\qquad
\nu_n = \frac{2 \pi}{\beta}\,n\qquad(n \in \mathbb{Z})
\label{1.6}
\end{align}
and translational invariance (in both space and time separately), it is easy to see 
that the spacetime volume factorizes,
%\footnote{The symbol $V$ denotes the volume of space.} 
$F(\mu_0) = - \ln Z(\beta) = \beta\,V\cdot f$. 
The free action density $f$ obtained in this way agrees, in the free case, with the 
thermodynamical free energy density of a non-interacting grand-canonical Bose gas 
(including the self-energy), provided that the path integral measure was properly 
normalized\footnote{This equality holds in units where $\hslash=1$ and the 
Boltzmann constant $k_B = 1$, which we assume from here on.} \cite{Bernard:1974bq}.
This observation  generalizes to interacting field theories, and in particular to gauge theories. 
In the latter case, the inclusion of the Faddeev-Popov (FP) determinant is crucial, 
and the Matsubara frequencies for the FP ghost fields must involve \emph{even} multiples of 
$\pi/\beta$ (as in eq.~(\ref{1.6})) even though the ghost fields obey Fermi statistics 
\cite{Bernard:1974bq}.

The inclusion of the ghost fields also requires an adaption of the variational principle: 
since the FP determinant $\JJ(A)$ describes the natural measure on the orbit space 
of the gauge fixed theory, the entropy of all trial measures must be computed \emph{relative}
to the FP determinant, which amounts to replacing the entropy by the relative entropy 
\cite{Quandt:2013wna}
\begin{align}
\Wb(\mu) &\equiv \WW(\mu) + \left\langle \ln( \JJ
) \right\rangle_\mu =- \left\langle \ln(\rho / \JJ
) \right\rangle_\mu \,.
\label{1.5}
\end{align}
 
\subsection{Gaussian trial measures}
\label{sec:2b}
The zero-temperature investigations in Ref.~\cite{Quandt:2013wna} have shown that 
a Gaussian ansatz for the path integral measure $d\mu$ is sufficient to describe 
the propagators of the theory accurately. We will therefore try a similar ansatz 
at finite temperatures,
\begin{align}
d\mu(A) = \NN\cdot dA\,\JJ(A)^{1-2\alpha}\cdot 
\exp\left[ - \frac{1}{2}\,\int_\beta d^4 (x,y)\,\, A_\mu^a(x)\,\delta^{ab}\,
\omega_{\mu\nu}(x,y)\,A_\nu(y)\right] \,,
\label{1.10}
\end{align}
where $\alpha$ is a variational parameter and the normalisation $\NN$ 
depends, in general, on $\alpha$ and on the kernel $\omega$. 
We have also assumed that the kernel $\omega$ can be taken colour diagonal because 
of global colour invariance left after (covariant) gauge fixing. 
Finally, we have also introduced the abbreviation
\begin{align}
\int_\beta dx \cdots \equiv \int_0^\beta dt\,\int_V d^3 x \cdots 
\end{align}
for the spacetime integral at finite temperature. The corresponding momentum integral is 
\begin{align}
\int_\beta \dd k\,f(\ka) \equiv  
\beta^{-1}\sum_{n \in \mathbb{Z}}\int \frac{d^3 k}{(2\pi)^3} \,f(\nu_n,\vek{k})
\cdots\,,
\label{1.12a}
\end{align}
i.e.~the integral over $k_0$ is always understood as a discrete sum ove the 
Matsubara frequencies $k_0 = \nu_n = 2 \pi n/\beta$.
The overall physical picture conveyed by the ansatz eq.~(\ref{1.10}) is a weakly 
interacting (constituent) gluon with an enhanced weight (for $\alpha > 0$) near 
the Gribov horizon.

\medskip\noindent
It is clear that the gauge field must be Fourier decomposed as in eq.~(\ref{1.6});
similarly, the FP determinant for covariant gauges
\begin{align}
\JJ(A) \equiv \Det \left[ - \partial_\mu\hat{D}_\mu^{ab} \right]
\,\big\slash\,\Det \left[ - \Box\,\delta^{ab}\right]
= \Det\,\left[ - \Box\,\delta^{ab} - g \,\partial_\mu f^{abc} A_\mu^c 
\right] \,\big\slash\, \Det \left[ - \Box\,\delta^{ab}\right]\, 
\label{1.11}
\end{align}
must be evaluated with periodic boundary conditions in time (even though the ghosts obey 
Fermi statistics).

\bigskip\noindent
Let us next look at the gluon propagator
\begin{align}
\big\langle\,A_\mu^a(x)\,A_{\nu}^b(y)\,\big\rangle = 
\int_\beta \dd k\,e^{ik(x-y)} \cdot\delta^{ab }\,D_{\mu\nu}(\ka)\,. 
\end{align}
At finite temperatures, $D_{\mu\nu}(\ka)$ is still a symmetric rank-2 tensor, but the overall 
$O(4)$ Lorentz invariance is broken because the heat bath singles out a rest frame specified by 
$u_\mu = (1,0,0,0)$. Assuming that the remaining \emph{spatial} $O(3)$ 
symmetry remains unbroken, $D_{\mu\nu}(\ka)$ must be a linear combination of all  
$O(3)$-invariant symmetric rank-2 tensors that can be formed from $k_\mu$ and $u_\mu$. 
Furthermore, BRST invariance entails the Ward identity for the longitudinal propagator,
\begin{align}
k^\mu k^\mu\,D_{\mu\nu}(\ka) = \zeta\,, 
\label{1.12}
\end{align}
where $\zeta$ is the coefficient for the gauge fixing term in covariant gauges,
$\mathcal{L}_{\rm fix}  = \frac{1}{2\zeta}\,\big(\partial_\mu A_\mu^a\big) ^2$. 
From all these constraints, the gluon propagator has just \emph{two}  remaining
Lorentz structures,\footnote{To simplify the notation, we will often write $f(\ka)$
instead of $f(\kargs)$ for a general function of the 4-momentum $k^\alpha = (\kargs)$.
If a function is $O(4)$ symmetric, $f(\ks)$ is really a scalar function of 
$k^2 \equiv k_0^2 + \vek{k}^2$; the distinction should always be clear 
from the context.} 
\begin{align}
D_{\mu\nu}(\ka) &= D_\perp(\ka)\,\PT_{\mu\nu}(\ka) + D_{\|}(\ka)\,\PL_{\mu\nu}(\ka) + 
\frac{\zeta} {k^2} \,\frac{k_\mu k_\nu}{k^2} 
\label{1.13} 
\\[2mm] 
D^{-1}_{\mu\nu}(\ka) &= D_\perp^{-1}(\ka)\,\PT_{\mu\nu}(\ka) + 
D_\|^{-1}(\ka)\,\PL_{\mu\nu}(\ka) + \frac{k_\mu k_\nu}{\zeta} \,.
\label{1.14}  
\end{align}
% The longitudinal pieces in the last term of both equations 
% receive no radiative corrections due to the Ward identity eq.~(\ref{1.12}). 
The two projectors $\PT$ and $\PL$ are both 4-dimensionally transversal, but $\PT$ is also 
3-dimensionally transversal, while $\PL$ is 3-dimensionally longitudinal, 
\begin{align} 
\PT_{\mu\nu}(\ka) &= (1-\delta_{\mu 0})\, (1-\delta_{\nu 0} )\,
\left( \delta_{\mu\nu} - \frac{k_\mu k_\nu}{\vek{k}^2} \right)
\label{1.15}
\\[2mm]
\PL_{\mu\nu} (\ka) &= \left(\delta_{\mu\nu}  - \frac{k_\mu k_\nu} {k^2} \right)
- \PT_{\mu\nu}(\kargs)\,.
\label{1.16}
\end{align}
These projectors enjoy the usual properties $(\PT )^2 = \PT$ and $(\PL) ^2 = \PL$, as 
well as $\PL\cdot \PT = \PT\cdot\PL = 0$; the traces in four euclidean dimensions are 
$\tr \,\PT = 2$ and $\tr\, \PL = 1$. Obviously, $\PT+\PL=\mathcal{P}$ is the 
usual 4-transversal projector.

\medskip\noindent
Returning to the Gaussian trial measure eq.~(\ref{1.10}), we note that, for 
$\alpha=\frac{1}{2}$, the kernel $\omega$ equals the inverse gluon propagator and is 
thus subject to a Lorentz decomposition similar to eq.~(\ref{1.14}). 
For $\alpha\neq\frac{1}{2}$, the FP determinant $\JJ$ contributes; below, we
will treat $\JJ$ in the so-called curvature approximation \cite{Reinhardt:2004mm} 
which maintains the Lorentz structure of the kernel. As a consequence, the kernel 
must have the general form for all $\alpha$,
\begin{align}
\omega_{\mu\nu}(\ka) = \wt(\ka) \cdot\PT_{\mu\nu}(\ka) 
+ \wl(\ka) \cdot\PL_{\mu\nu}(\ka) + \gamma^{-1} \,k_\mu k_\nu\,,
\label{1.20}
\end{align}
where $\gamma =\zeta$ up to radiative corrections. 
The two scalar dressing functions $\wt$ and $\wl$ depend on $k_\mu$ 
only through the two invariants $k_0$ and $|\vek{k}|$. 
% For later reference, we note that the full $O(4)$ invariance in eq.~(\ref{1.20}) 
% is recovered if $\wt=\wl$ depends on $k^2$ only.
  
\subsection{Curvature approximation and the ghost DSE}
\label{sec:2c}
To complete our computational tools, we must also evaluate the FP determinant 
$\JJ(A)$  from eq.~(\ref{1.11}) and its expectation value in the 
trial measure (\ref{1.10}). Unfortunately, this cannot be computed in closed form
and we will resort to the finite-temperature equivalent of the so-called \emph{curvature 
approximation} which was previously shown to be effective both in the Hamiltonian approach 
to Coulomb gauge \cite{Reinhardt:2004mm}, and in covariant gauges at zero temepratures 
\cite{Quandt:2013wna}. In this approximation, the FP determinant is expressed as
%\footnote{This approximation is exact up to two loop, if a formal loop 
%counting  parameter is introduced in the exponent of eq.~(\ref{1.10}).} 
\begin{align}
\ln \JJ[A]\approx - \frac{1}{2}\int_\beta d(x,y)\,\chi^{ab}_{\mu\nu}(x,y)
\cdot A_\mu^a(x)\,A_\nu^b(y)\,,
\label{15} 
\end{align}
where the \emph{curvature} is given by the expectation value
\begin{align}
 \chi^{ab}_{\mu\nu}(x,y) \equiv 
-\,\left\langle \frac{\delta^2 \ln \JJ}{\delta A_\mu^a(x)\,\delta A_\nu^b(y)}
\right \rangle
\label{16}
\end{align}
taken with the Gaussian measure eq.~(\ref{1.10}). From global colour invariance, 
we have again $\chi_{\mu\nu}^{ab} = \delta^{ab}\,\chi_{\mu\nu}$. 
Furthermore, the arguments put forward above to determine the Lorentz structure 
of the propagator also apply to the curvature. As a consequence, we have 
\begin{align}
\chi_{\mu\nu}(\ka) = \xt(\ka)\cdot\PT_{\mu\nu}(\ka) + \xl(\ka)\cdot\PL_{\mu\nu}(\ka) 
+ \cdots\,,
%+ k_\mu k_\nu\,\tau(\ka)\,.
\label{1.17}
\end{align}
where the dots indicate irrelevant 4-longitudinal pieces which vanish in 
Landau gauge. We will derive integral equations for the profile 
functions in the next section.

\bigskip\noindent
If we put everything together, our \emph{Ansatz} (\ref{1.10}) for the trial measure becomes
\begin{align}
d\mu(A) = \NN\cdot \exp\Bigg[ - \frac{1}{2}\,\int_\beta \dd k\, 
A_\mu^a(-k)\,\Big \{&\wb(\ka) \cdot \PT_{\mu\nu}(\ka)  + \sb(\ka)\cdot \PL_{\mu\nu}(\ka)
+\tb(\ka)\cdot k_\mu k_\nu \Big\}\Bigg ]\,,
\label{1.18}
\end{align}
where the profile functions are now
\begin{align} 
\wb &\equiv \wt + (1-2\alpha)\,\xt \nonumber\\
\sb &\equiv \wl + (1-2\alpha)\,\xl\,. 
\label{1.19}
\end{align}
A few remarks are in order at this point. Firstly, the full gluon propagator in our 
approach is
\begin{align}
D_{\mu\nu}^{ab}(\ka) = \delta^{ab}\,\left[\wb(\ka)^{-1}\cdot \PT_{\mu\nu}(\ka) + 
\sb( k)^{-1}\cdot\PL_{\mu\nu}(\ka) + \tb(\ka)^{ -1}\cdot\frac{k_\mu k_\nu}{k^4}
\right]\,.
\label{1.22}
\end{align}
The $4$-longitudinal (last) piece in this expression is not very important and 
will even vanish in Landau gauge $\zeta\to 0$, which we will study exclusively below.

Secondly, the operator held fixed in our trial measure is not the gauge field itself, 
but rather the gluon propagator (cf.~eq.~(\ref{1.22})), i.e.~according to the 
general setup explained in section \ref{sec:2a}, our variation principle 
optimizes the \emph{effective action for the gluon propagator}. This is appropriate 
as long as we are not interested in higher vertices.

Finally, one might think that the curvature is irrelevant, as the final form of 
the trial measure including the FP determinant is still Gaussian. This is not so
because the curvature reappears in the relative entropy $\Wb$ that enters the free 
action. 

%%%%%%%%%%%%%%%%%%%%%%%%%%%%%%%%%%%%%%%%%%%%%%%%%%%%%%%%%%%%%%%%%%%%%%%%%%%%%%%%%%%%%%%%

\subsection{The ghost sector}
\label{sec:2d}
The ghost propagator is the expectation value of the inverse Faddeev-Popov operator
which we decompose according to $G^{-1} \equiv - \partial_\mu\,\hat{D}_\mu = G_0^{-1} + h$, 
where
\begin{align}
\big[ G_0^{-1}\big]^{ab} = - \Box\,\delta^{ab} \,,\qquad\qquad h^{ab}  = 
g\,f^{abc}\,\partial_\mu A_\mu^c
\end{align}
with the structure constants $f^{abc}$ of the colour group $SU(N)$. From this, it is 
easy to see that the ghost propgator $\langle G \rangle$ satisfies the Dyson equation
$
\langle G\rangle^{-1} = G_0^{-1} + \langle h G\rangle\,\langle G \rangle^{-1}\,.
$
The expectation value $\langle h G\rangle$ involves the ghost-gluon Green's function, 
which in turn is obtained by attaching full propagators to the proper ghost-gluon vertex 
$\Gamma^{abc} _\mu$. Using roman digits for the combination of Lorentz, colour and 
space-time indices, the  ghost gluon vertex is defined by
\begin{align}
\big\langle\,A(1)\,G(2,3)\,\big\rangle = - 
D(1,1')\,\big\langle\, G(2,2')\,\big\rangle\,\Gamma(2',3'\;;\,1')\,
\big\langle\,G(3'3)\,\big\rangle\,.
\label{400c}
\end{align}

If we also introduce the \emph{ghost form factor}  $\eta(\ka)$ as the 
deviation of the full ghost propagator from the free one, 
\begin{align}
\big\langle G\big\rangle^{ab}(x,y) = \int_\beta \dd k\,e^{ik\cdot (x-y)} \,
\delta^{ab}\,\frac{\eta(\ka)}{k^2}\,,
\label{400} 
\end{align}
we find the \emph{exact} momentum space relation
\begin{align}
\delta^{ab}\,\eta(\ka)^{-1} = \delta^{ab} + g\,f^{acd} \int_\beta \dd q\,
\frac{i k_\mu} {k^2} \,D_{\mu\nu}(\qa)\,\frac{\eta(k-q)}{(k-q)^2}\,
\Gamma^{dcb}_\mu(q-k,k) \,.
\label{401}
\end{align}
Here, we have used global colour invariance to deduce the colour structure
$D^{ab}_{\mu\nu} = \delta^{ab} D_{\mu\nu}$ for the full gluon propagator, and
the momentum routing in the vertex is
\begin{align*}
\Gamma^{dcb }_\mu(q-k,k) \,: \qquad\quad
\begin{minipage}[c]{5cm} \includegraphics[width=4.5cm]{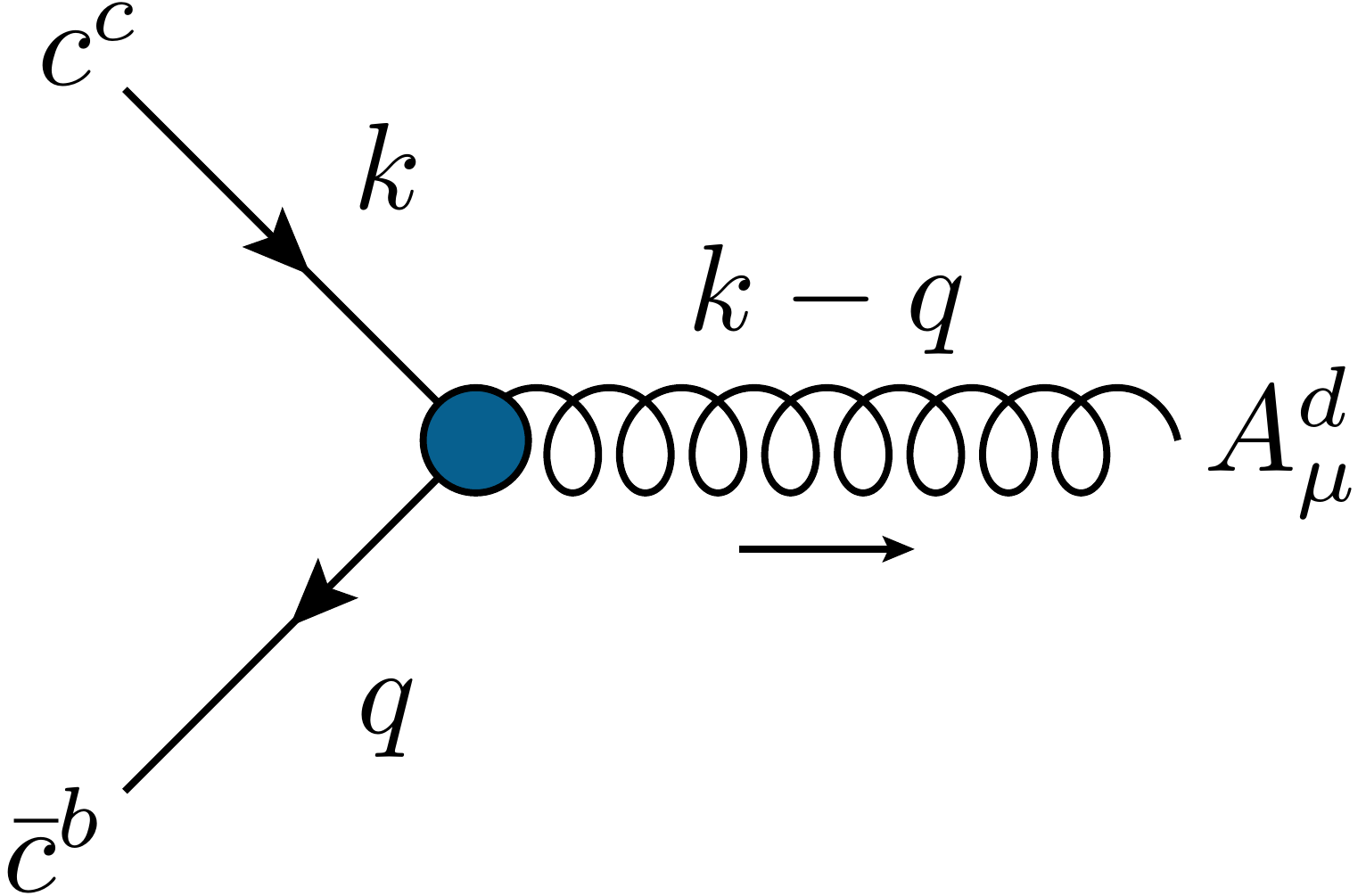}\end{minipage} 
\end{align*}
The exact eq.~(\ref{401}) is referred to as the ghost Dyson-Schwinger equation (DSE).
In the \emph{rainbow-ladder approximation}, the proper ghost-gluon vertex is replaced 
by the bare vertex defined by
\begin{align}
\Gamma_0(2,3\,;\,1) = \frac{\delta G^{-1}(2,3)}{\delta A(1)}\,,
\label{G9}
\end{align}
which reads, with the momentum routing above,
\begin{align} 
\big[\Gamma_0\big]^{dcb}_\mu(q-k,k) \to i g \,f^{dcb}\,(q-k)_\mu\,. 
\label{402}
\end{align}
The ladder approximation has been proven to be very reliable at $T=0$ and we expect 
it to hold equally well at non-zero temperatures. From eq.~(\ref{401}), we then get a 
closed integral equation for the ghost form factor, 
\begin{align} 
\eta(\ka)^{-1} = 1 &- Ng^2\,\int_\beta\dd q\,\frac{\eta(k-q)}{(k-q)^2 }\,
\frac{k_\mu (k-q)_\nu\,D_{\mu\nu}(\qa)}{k^2}\,.
\label{403}
\end{align}
To complete the derivation of the ghost sector, we have to insert the gluon propagator 
eq.~(\ref{1.20}) from our Gaussian ansatz into eq.~(\ref{403}). 
This simplifies considerably in pure Landau gauge ($\zeta=0$) when the gauge connection and the 
gluon propagator are 4-dimensionally transversal. From this point on, \emph{we will therefore study 
Landau gauge ($\zeta=0$) exclusively}. In this case, the ghost DSE (\ref{403}) becomes
\begin{align} 
\eta(\ka)^{-1} = 1 - Ng^2\,\int_\beta\dd q\,&\frac{\eta(k-q)}{(k-q)^2 }\,
\frac{1 - (\unitfour{k}\cdot\unitfour{q})^2}{\sb(\qa)} - 
\nonumber\\[2mm]
&-Ng^2\cdot\frac{\vek{k}^2}{k^2}\int_\beta\dd q \,\frac{\eta(k-q)}{(k-q)^2 }\,
\big(1 - (\unitthree{k}\cdot\unitthree{q})^2\big)\,
\Big [\wb(\qa)^{-1} - \sb(\qa)^{-1}\Big] \,,
\label{480}
\end{align}
where $k^2 = k_0^2 + \vek{k}^2$ is the 4-momentum square and we have also introduced 
four- and three-dimensional unit vectors
\begin{align}
\unitfour{k}_\mu = \frac{k_\mu}{\sqrt{k^2}}\,,\qquad\qquad\qquad
\unitthree{k} = \frac{\vek{k}}{|\vek{k}|}\,.
\end{align}
 
\medskip\noindent
Next, we concentrate on the curvature eq.~(\ref{16}). In our compact roman digits
notation, we have
\begin{align*}
\chi(1,2) = \mathrm{Tr}\cdot\left\langle G\,\frac{\delta(-\partial \hat{D})}{\delta A(2)}\cdot
G\cdot \frac{\delta(-\partial \hat{D})}{\delta A(1)}\right\rangle
\end{align*}
The derivatives yield the bare ghost-gluon vertex $\Gamma_0$ (\ref{G9}), which is 
field-independent. To the given (formal) loop order, we therefore have
$
\langle G\,\Gamma_0\,G\rangle \approx \langle G \rangle\,\Gamma_0\,\langle G \rangle\,. 
$
The remaining expectation values are merely the ghost propagators computed earlier. 
Restoring all arguments, we obtain the momentum space expression 
\begin{align}
\chi^{ab}_{\mu\nu}(\ka) = - \delta^{ab}\,Ng^2\int_\beta \dd q\,
\frac{\eta(k-q)\,\eta(\qa)}{(k-q)^2\,q^2}\,(k-q)_\mu\,q_\nu\,. 
\label{404}
\end{align}
In Landau gauge, only the 4-transversal parts of this tensor contribute, 
which we find by contracting with the corresponding projectors $\PT$ and $\PL$, respectively. 
After some straightforward algebra, the profile functions in eq.~(\ref{1.17}) are
\begin{align}
\xt(\ka) &= \frac{1} {2}\,Ng^2\int_\beta \dd q\,\frac{\eta(k-q)\,\eta(\qa)}{(k-q)^2}\,
\frac{\vek{q}^2}{q^2}\,\Big[ 1 - \big(\unitthree{k}\cdot\unitthree{q}\big)^2 \Big] 
\nonumber\\[3mm]
\xl(\ka) &= Ng^2\,\int_\beta \dd q\,\frac{\eta(k-q)\,\eta(\qa)}{(k-q)^2}\,
\left\{ 1 - \big(\unitfour{k}\cdot\unitfour{q}\big)^2 - 
\frac{\vek{q}^2} {q^2}\,\Big[1 - \big(\unitthree{k} \cdot\unitthree{q}\big)^2\Big] 
\right\}\,.
\label{405}
\end{align}

%%%%%%%%%%%%%%%%%%%%%%%%%%%%%%%%%%%%%%%%%%%%%%%%%%%%%%%%%%%%%%%%%%%%%%%%%%%%%%%%%%%%%%%%

\section{The gap equation at finite temperatures}
\label{sec:3}
\subsection{The classical action}
\label{sec:3a}
To derive the effective action for the gluon propgator, we  must first evaluate the 
expectation value of the gauge-fixed YM action in the trial measure eq.~(\ref{1.10}),
\begin{align}
\big\langle\, S_{\mathrm{gf}}\,\big\rangle &= 
\frac{1}{2} \int_\beta d x\,\big\langle\,A_\mu^a(x)\,A_\nu^a(y)\,\big\rangle\,
\big[-\Box_x\,\delta_{\mu\nu} + (1- \xi^{-1})\,\partial_\mu^x\,\partial_\nu^x\big]\,\delta(x,y) +
\label{28} \\[2mm]
&+ g\,f^{abc}\,\int_\beta d x\,\partial_\mu^x\,\big\langle\,A_\nu^a(x)\,A_\mu^b(x)\,A_\nu^c(x)\,
\big\rangle
+ \frac{g^2}{4}\,f^{abc}\,f^{ade}\,\int_\beta d x \,\big\langle\,A_\mu^b(x) A_\nu^c(x)
A_\mu^d(x)A_\nu^e(x)\,\big\rangle \,.
\nonumber
\end{align}
The relevant correlators are easily evaluated using Wick's theorem, global colour 
invariance, and the $SU(N)$ relation $f^{abc} f^{abc} = N(N^2-1)$ to perform the 
colour traces.  In momentum space, the result takes the form
\begin{align}
\big\langle\, S_{\mathrm{gf}}\,\big\rangle = &
\frac{1}{2}\,(N^2-1)\,\beta V\int_\beta \dd k\,\Big(k^2\,\delta_{\mu\nu}-\big(1-\zeta ^{-1}\big)\,
k_\mu k_\nu\big)\,D_{\mu\nu}(\ka) 
\\[2mm] 
&{}+ \frac{Ng^2}{4}\,(N^2-1)\,\beta V\,\left[\int_\beta \dd k \,D_{\mu\mu}(\ka)\right] ^2
-\frac{Ng^2}{4}\,(N^2-1)\,\beta V\int_\beta \dd(k,q) D_{\mu\nu }(\ka)\,D_{\nu\mu}(\qa)\,,
\nonumber
\end{align}
where $V$ is the 3-dimensional space volume. Next, we insert the representation 
(\ref{1.22}) and take the Landau gauge limit $\zeta\to 0$ to 
find\footnote{We have dropped the field- and temperature-independent constant
\[\frac{1}{2}\,\beta V\,(N^2-1)\,\int_\beta \dd k
= \frac{V}{2}\,(N^2-1)\,\sum_{n \in \mathbb{Z}}\int\frac{d^3k}{(2\pi)^3}\,.\]}
\begin{align}
\frac{\big\langle\, S_{\mathrm{gf}}\,\big\rangle}{(N^2-1)\,\beta V} = &
\frac{1}{2}\int_\beta \dd k\,k^2\left[\frac{d-1}{\wb(\ka)} + \frac{1}{\sb(\ka)}\right]
+ \frac{Ng^2}{4}\,\left\{\int_\beta \dd k \left[ \frac{d-1}{\wb(\ka)} + \frac{1}{\sb(\ka)}
\right] \right\}^2
\nonumber\\[3mm]
&-\frac{Ng^2}{4}\,\int_\beta \dd(k,q) \Bigg\{
\frac{1+(\unitthree{k}\cdot\unitthree{q}) ^2 }{\wb(\ka)\,\wb(\qa)} + 
\frac{2}{\wb(\ka)\sb(\qa)}\,\Big(d-2-(\unitthree{k}\cdot\unitthree{q})^2\Big)\,
\left(1-\frac{\vek{q}^2}{q^2}\right)
\nonumber \\[2mm]  
&\qquad\quad 
+ \frac{1}{\sb(\ka)\, \sb(q )}\,\left[\big(\unitfour{k}\cdot\unitfour{q}\big)^2 + 
\Big(1 - (\unitthree{k}\cdot\unitthree{q})^2\Big)\,\left(-1 +  
\frac{\vek{k}^2}{k^2 } + \frac{\vek{q}^2}{q^2 } \right) \right]\Bigg\}\,.
\label{408}
\end{align}
To further simplify this expression, we can exploit the remaining rotational symmetry 
to rewrite some of the integrals in which the integrand depends on $O(3)$ invariants only.
%\begin{align}
%\int_\beta \dd q \,f(q_0^2,\vek{q}^2) \,(\unitthree{k}\cdot\unitthree{q})^2 = 
%\frac{1}{3}\,\int_\beta \dd q\,f(q_0^2,\vek{q}^2) \,. 
%\end{align}
After some straightforward but lengthy algebra, the classical action can be recast to
\begin{align}
\big\langle\, S_{\mathrm{gf}}\,\big\rangle = &
\frac{1}{2}\,(N^2-1)\,\beta V\int_\beta \dd q\,q^2\left[\frac{2}{\wb(\qa)} + \frac{1}{\sb(\qa)}\right]
\nonumber\\[2mm]
&+ \frac{1}{4}\,Ng^2\,(N^2-1)\,\beta V\,\int_\beta \dd(k,q)\Bigg\{
\frac{A}{\wb(\ka)\,\wb(\qa)} + \frac{2 B(\qa)}{\wb(\ka) \,\sb(\qa)} + \frac{C(\ka,\qa)}{\sb(\ka)\,\sb(\qa)}
\Bigg\}\,, 
\label{406}
\end{align}
where the coefficients are given, for $d=3$ space dimensions, by
\begin{align}
A &= \frac{8}{3}
\nonumber\\
B(\qa) &= \frac{4}{3}\,+ \frac{2}{3} \,\frac{\vek{q}^2}{q^2}
\nonumber \\ 
C(\ka,\qa) &= \frac{2}{3} + \frac{1}{3}\,\left(\frac{\vek{k}^2}{k^2}  + \frac{\vek{q}^2}{q^2}\right)
- \frac{4}{3}\,\frac{\vek{k}^2}{k^2}\,\frac{\vek{q}^2} {q^2}
\,.
\label{410}
\end{align}

\subsection{The entropy}
\label{sec:3b}
As the last ingredient, we need the relative entropy of the path integral measure 
(\ref{1.18}) with respect to the Faddeev-Popov determinant. 
From eq.~(\ref{1.5}) and the curvature approximation eq.~(\ref{15}), we obtain 
\begin{align*}
\Wb &= \big \langle - \ln \rho \big \rangle + \big \langle \ln \JJ \big \rangle
\\[2mm]
&\approx - \ln \NN + \frac{1}{2}\,\int_\beta d (x,y)\, \bar{\omega}^{ab}_{\mu\nu}(x,y)\,
\big\langle A^a_\mu(x)\,A_\nu^b(y)\big \rangle -
\frac{1}{2}  \,\int_\beta d (x,y)\, \chi^{ab}_{\mu\nu}(x,y)\,
\big\langle A^a_\mu(x)\,A_\nu^b(y)\big \rangle 
\\[2mm]
&= - \ln \NN + \frac{1}{2}\int_\beta d(x,y)\,
\langle A_\mu^a(x)A_\nu^b(y)\rangle
\,\Big\{ \bar{\omega}^{ab}_{\mu\nu}(x,y)-\chi^{ab}_{\mu\nu}(x,y)\Big\}
\\[2mm]
&= - \ln \mathrm{ det}\left(\frac{2\pi}{\bar{\omega}}\right)^{-\frac{1}{2}}
+ \frac{1}{2}\,(N^2-1)\int_\beta d(x,y)\,
\bar{\omega}^{-1}_{\mu\nu}(x,y) \,\Big\{ \bar{\omega}_{\mu\nu}(x,y)-\chi_{\mu\nu}(x,y)\Big\}
\\[2mm]
&=
-\frac{1}{2}\,\mathrm{Tr}\ln\left( \frac{\bar{\omega}^{ab}_{\mu\nu}}{2\pi}\right) + 
\frac{1}{2}\,(N^2-1)\,\beta V\int_\beta \dd k\,\bar{\omega}^{-1}_{\mu\nu}(\ka)\,
\Big\{ \bar{\omega}_{\mu\nu}(\ka)-\chi_{\mu\nu}(\ka)\Big\}\,,
\end{align*}
where the explicit form $\NN = \det[\bar{\omega}/(2\pi)]^\frac{1}{2}$ for the 
normalisation in eq.~(\ref{1.18}) was used. In pure Landau gauge in 
three space dimensions, we have 
$\bar{\omega}^{-1}_{\mu\nu}(\ka)\,\bar{\omega}_{\mu\nu}(\ka) = \tr\,\mathcal{P}(\ka) = 3$, 
where $\mathcal{P}$ is the 4-transversal projector. Likewise, 
\begin{align*}
\bar{\omega}^{-1}_{\mu\nu}(\ka)\,\chi_{\mu\nu}(\ka)
&= \wb^{-1}(\ka)\,\xt(\ka)\cdot \tr\,\PT(\ka) + \sb^{-1}(\ka)\,\xl(\ka)
\cdot \tr\,\PL(\ka)
\\
&=2\,\wb(\ka)^{-1}\,\xt(\ka) + \sb(\ka)^{-1}\,\xl(\ka)\,.
\end{align*}
After dropping an irrelevant temperature- and field-independent term, 
the entropy becomes
\begin{align*}
\Wb &= - \frac{1}{2}\,(N^2-1)\,\mathrm{Tr}\ln\,\left[
\frac{\wb}{2\pi}\,\PT_{\mu\nu} + \frac{\sb}{2\pi}\,
\PL_{\mu\nu}\right] - \frac{1}{2}\,(N^2-1)\,\beta V\int_\beta
\dd k\,\Big\{ 2\,\frac{\xt(\ka)}{\wb(\ka)} + \frac{\xl(\ka)}{\sb(\ka)}
\Big\}\,.
\end{align*}
Since the projectors $\PT$ and $\PL$ are orthogonal, the determinant in 
the first term factorizes. The dimensions of the subspaces on which 
$\PT$ and $\PL$ project are $2$ and $1$, respectively, and in each subspace, 
the projector acts as unity. Expressing the functional trace in momentum space,
we finally obtain
\begin{align}
\Wb &= - \frac{1}{2}\,(N^2-1)\,\beta V\int_\beta \dd k 
\Bigg\{
2\,\ln\frac{\wb(\ka)}{2\pi} + \ln\frac{\sb(\ka)}{2\pi} + 
2 \,\frac{\chi(\ka)}{\wb(\ka)} + \frac{\theta(\ka)}{\sb(\ka)}
\Bigg\}\,.
\label{412}
\end{align}
The difference of eqs.~(\ref{406}) and (\ref{412}) is the effective action 
$\Gamma(\wb,\sb)$ for the (inverse) gluon propagator within 
the curvature approximation (\ref{15}).

\subsection{The gap equation}
\label{sec:3c}
It is now straightforward to compute the optimal kernels (or gluon propagator) 
from the \emph{gap equations} 
\begin{align}
\frac{\delta \Gamma(\wb,\sb)}{\delta \wb(\ka)} = 
\frac{\delta \Gamma(\wb,\sb)}{\delta\sb(\ka)} = 0\,. 
\end{align}
After more algebra, we obtain for the variation w.r.t~$\wb(\ka)$
\begin{align}
0 = k^2 + \frac{1}{2}\,Ng^2\,\int_\beta \dd q\left[
\frac{A}{\wb(\qa)} + \frac{B(\qa)}{\wb(\qa)}\right] - \wb(\ka) + \chi_\perp(\ka) + 
\int_\beta \dd q\,\frac{1}{\wb(\qa)}\,\frac{\delta \chi(\qa)}
{\delta \wb(\ka)^{-1}}\,,
\label{418}
\end{align}
where the coefficients $A$ and $B(\qa)$ are given in eq.~(\ref{410}).
The curvature depends on $\wb(\qa)$ through the ghost form factor $\eta$
which enters eq.~(\ref{405}), but this is a higher loop effect that can safely 
be neglected within the curvature approximation. Dropping the last term in 
eq.~(\ref{418}), we arrive at
\begin{align}
\wb(\ka) &= k_0^2 + \vek{k}^2 + \xt(\ka) + \Mt^2(\beta)
\label{420}
\\[2mm]
\Mt^2(\beta) &\equiv \frac{1}{2}\,Ng^2\,\int_\beta \dd q\,\left[\frac{A}{\wb(\qa)}
+ \frac{B(\qa)}{\sb(\qa)}\right]\,.
\label{422}
\end{align}
The second gap equation is derived in exactly the same fashion and takes the form
\begin{align}
\sb(\ka) = k_0^2 + \vek{k}^2 + \xl(\ka) + Ng^2\,\int_\beta \dd q
\left[ \frac{B(\ka)}{\wb(\qa)} + \frac{C(\ka,\qa)}{\sb(\qa)}\right]\,.
\label{424}
\end{align}
In constrast to the previous case (\ref{422}), the last piece has no direct 
interpretation as a mass term as it depends on the external momentum $k_\mu$. 
However, at $\vek{k}=0$ (with $k_0 \neq 0$), we find
\begin{align*} 
Ng^2\,\int_\beta \dd q\,\left[\frac{B(\ka)}{\wb(\qa)} + 
\frac{C(\ka,\qa)}{\wb(\qa)}\right] \stackrel{\vek{k}=0}{\longrightarrow} 
\Mt^2(\beta)\,,
\end{align*}
as can be shown by explicit calculation. We can thus factorize the momentum 
dependence in the last term on the rhs of eq.~(\ref{424}) and recast the 
second gap equation to
\begin{align}
\sb(\ka) &= k_0^2 + \vek{k}^2 + \xl(\ka) + \Mt^2(\beta) + \frac{\vek{k}^2}
{k_0^2 + \vek{k}^2}\, \Ml^2(\beta)
\label{430}
\\[2mm]
\Ml^2(\beta) &= \frac{1}{3}\,Ng^2\,\int_\beta \dd q
\left[ \frac{2}{\wb(\qa)} + \left(\frac{q_0^2 - 3 \vek{q}^2}{q_0^2
+ \vek{q}^2}\right)\,\frac{1}{\sb(\qa)}\right]\,.
\label{432}
\end{align}
Eqs.~(\ref{403}), (\ref{405}), (\ref{420}) and (\ref{430}) form a closed set 
of integral equations to determine the gluon and ghost propagators in Landau gauge 
at finite temperatures.

\section{Renormalization}
\label{sec:4}
Next we turn to the renormalization of our integral equation system. We expect that 
\textbf{(i)} the equations reduce to the known zero-temperature case 
in the limit $\beta \to \infty$ and \textbf{(ii)} the counterterms fixed in the 
zero-temperature limit are sufficient to renormalize the system at \emph{any} temperature.

\subsection{The zero temperature limit}
\label{sec:4a}
Let us first study how the zero temperature limit formally arises in our 
integral equation system.\footnote{Since we are dealing with the unrenormalized equations, 
an $O(3)$ invariant cutoff is implicitly assumed in this section.} The sum over Matsubara
frequencies can generally be rewritten by Poisson resummation,
\begin{align}
 \frac{1}{\beta}\sum_{n=-\infty}^\infty f(\nu_n) = \int_{-\infty}^\infty 
 \frac{dz}{2\pi}\,f(z)\sum_{m=-\infty}^\infty e^{i m \beta z}
= \sum_{m=-\infty}^\infty \tilde{f}(m \beta)\,.
\label{X1}
\end{align}
By the Riemann-Lebesgue lemma, the Fourier transform $\tilde{f}(x)$ vanishes at large 
arguments $|x|\to \infty$, i.e.~only the term with $m=0$ contributes in the zero-temperature
limit,
\begin{align}
\lim_{\beta \to \infty}   \frac{1}{\beta}\sum_{n=-\infty}^\infty f(\nu_n) = 
\tilde{f}(0) = \int \frac{dp_0}{2\pi} \,f(p_0)\,
\label{X23}
\end{align}
Let us assume for the moment that the ghost form factor $\eta(\ka)$ approaches the 
$O(4)$-invariant zero-temperature form factor $\eta_0(\ks)$, cf.~appendix \ref{app:zero}. 
(This assumption will be justified \emph{a posteriori} below). From eq.~(\ref{404})
and the limit (\ref{X23}), we have  
\begin{align*}
\chi^{ab}_{\mu\nu}(\ka) \stackrel{\beta\to\infty}{\longrightarrow} - \delta^{ab}\,Ng^2
\int \frac{d^4 q}{(2\pi)^4}\,\frac{\eta_0(k-q)\,\eta_0(\qs)}{(k-q)^2\,q^2}\,(k-q)_\mu\,q_\nu\,.
\label{503}
\end{align*}
Contracting with $\PT_{\mu\nu}(\ka)$ gives
\[
\lim_{\beta\to\infty} 2 \xt(\ka) = -  \delta^{ab}\,Ng^2
\int \frac{d^4 q}{(2\pi)^4}\,\frac{\eta_0(k-q)\,\eta_0(\qs)}{(k-q)^2\,q^2}\,(k-q)_\mu\,q_\nu\,
\PT_{\mu\nu}(\ka) = 2 \chi_0(\ks)\,,
\]
where $\chi_0$ is the zero-temperature curvature, cf.~appendix \ref{app:zero}.
From the second equation (\ref{405}), it follows that $\xl(\ka) \to 3\,\chi_0(\ks)- 
2 \lim\limits_{\beta\to\infty}\xt(\ka) = \chi_0(\ks)$ in the same limit and we have
\begin{align}
\lim_{\beta\to\infty} \xt(\ka)=  \lim_{\beta\to\infty} \xl(\ka)= \chi_0(\ks)\,. 
\label{504}
\end{align}
In appendix \ref{app:mass}, it is further shown that the mass functions in the gap equation 
have the low-temperature limit $\Mt^2(\beta) \to M_0^2$ and $\Ml^2(\beta) \to 0$. 
From eqs.~(\ref{420}), (\ref{430}) and (\ref{504}), it then follows immediately that
\begin{align}
\lim_{\beta\to\infty} \wb(\ka)=  \lim_{\beta\to\infty} \sb(\ka)= 
k_0^2 + \vek{k}^2 + \chi_0(\ks) + M_0^2 = \bar{\omega}_0(\ks)\,.
\label{506}
\end{align}
Finally, this relation shows that the second line in eq.~(\ref{480}) vanishes 
as $\beta\to\infty$, while the first line approaches $\eta_0(\ka)$ as can be seen 
from eqs.~(\ref{501}) and (\ref{502}). Thus, the self-consistent solution has also
\begin{align}
\lim_{\beta\to\infty} \eta(\ka)  = \eta_0(\ks)\,,  
\label{508}
\end{align}
which justifies our initial assumption above. Alltogether, eqs.~(\ref{504})-(\ref{508})
are the expected $\beta\to\infty$ limit in which the full $O(4)$ symmetry is restored.

\subsection{Renormalization at zero temperature}
\label{sec:4b}
In this section, we recall (and simplify) the renormalization of the zero temperature 
system as layed out in Ref.~\cite{Quandt:2013wna}. This information is required
later on, since we must fix all counter terms at zero temperature in order to compare different
temperature settings reliably.
After introducing ghost fields $c,\bar{c}$ to make the Faddeev-Popov determinant local, 
three counter terms are required,
\begin{align}
\mathcal{L}_{\mathrm{ct}} = \delta Z_A\cdot \frac{1}{4} \,(\partial_\mu A_\nu^a - 
\partial_\nu A_\mu^a)^2 + \delta M^2\cdot \frac{1}{2}\,\big(A_\mu^a\big)^2
+ \delta Z_c\cdot \partial_\mu \bar{c}\,\partial^\mu c
\label{510}
\end{align}
corresponding to a gluon wave function, gluon mass, and ghost field 
renormalisation. (No vertex renormalisation is induced by the theory.)
The counterterms $\int d^4x\,\mathcal{L}_{\mathrm{ct}}$
must be added to the exponent of the (Gaussian) trial measure eq.~(\ref{1.10})
used to compute the $n$-point functions, i.e.~they are \emph{not} counterterms 
to the Yang-Mills action directly. The renormalized integral equation system 
at zero temperature then becomes
\begin{alignat}{3}
\bar{\omega}_0(\ks) &= k^2 + M_0^2 + \chi_0(\ks) & \qquad & \qquad &
M_0^2 &= N g^2\,I_M^{(0)} + \delta M_1^2 
\nonumber \\[2mm]
\eta_0(\ks)^{-1} &= 1 - N g^2\,I_\eta^{(0)}(k) - \delta Z_c & \qquad & \qquad &
\chi_0(\ks) &= N g^2\,I_\chi^{(0)}(k) + \delta \chi + k^2\,\delta Z_A\,,
\label{512}
\end{alignat}
where the explicit form of the various loop integrals is listed in appendix \ref{app:zero}. 
The mass counterterm has been split, $\delta M^2 = \delta M_1^2 + 
\delta \chi$, between the mass and the curvature since only the sum of these
contributions enters the gap equation (\ref{512}). 

As layed out in Ref.~\cite{Quandt:2013wna}, we have to choose slightly unusual renormalization conditions 
to overcome numerical problems in the deep infrared. Normally, one would prescribe the value of the 
propagators and vertices at a common (large) scale $\mu \gg 1$, except for the mass counterterm which 
determines the propagator at zero momentum (in Euclidean space). In the present case, we impose the 
value of the \emph{ghost propagator}, or ghost form factor $\eta(\mu_c)$, at a low scale $\mu_c \ll 1$ 
(which could even be taken to be $\mu_c=0$) in order to discern  the scaling from the decoupling 
solution \cite{Quandt:2013wna}; this condition determines the counter term
\begin{align}
\delta Z_c =  1 - Ng^2\,I_\eta^{(0)}(\mu_c) - \eta_0(\mu_c)^{-1} \,.
\label{513}
\end{align}
The remaining renormalization conditions for the gluon can be taken conventionally: we impose 
the value of the (inverse) propagator at a large scale $\mu \gg 1$ to fix the field renormalisation 
$\delta Z_A$, and at a small scale $\mu_0 \ll 1$ to fix the mass counterterm\footnote{For the decoupling 
solution, we could choose $\mu_0 \to 0$ so that $M_A^2$ could be interpreted as a constituent gluon mass; 
in the general case, however, we must take $\mu_0 > 0$ to avoid infrared singularities, and $M_A^2$ 
becomes a general mass parameter.},
\begin{align}
\bar{\omega}_0(\mu) =: Z\,\mu^2
\,,\qquad\qquad\quad 
\bar{\omega}_0(\mu_0) =: Z\,M_A^2
\,.
\label{524}
\end{align}
For later reference, we note the explicit form of the counterterm 
coefficients\footnote{The symbol $Z$ should not be confused with the gluon field renormalization 
constant, which is $Z_A = 1 + \delta Z_A$. 
%The gluon propagator is $D(k)  = Z_A^{-1}\,(1 + \Delta R) / k^2 \approx (1 + \Delta R - \delta Z_A) / k^2$, 
%where $\Delta R$ are loop corrections, and $\delta Z_A = \Delta R\big\vert_\infty + \text{finite}$. 
%The same conclusion follows from the requirement that the gluon propagator for the bare field 
%$A^{(B)} = Z_A^{\frac{1}{2}}\,A$ has unit residue.
%In the absence of interactions, $M_A = 0$ and $I_\chi^{(0)} = 0$ so that $Z_A = Z$.
}
\begin{align}
1 + \delta Z_A &= Z\,\frac{\mu^2 - M_A^2}{\mu^2-\mu_0^2} - Ng^2\,\frac{I_\chi^{(0)}(\mu) - I_\chi^{(0)}(\mu_0)}
{\mu^2-\mu_0^2}
\nonumber\\[2mm]
M_0^2 + \delta M^2 &= Z\,\mu^2\,\frac{M_A^2 - \mu_0^2}{\mu^2-\mu_0^2} - Ng^2\,
\frac{\mu^2\,I_\chi^{(0)}(\mu_0) - \mu_0^2\,I_\chi^{(0)}(\mu)}{\mu^2-\mu_0^2}\,,
\label{530}
\end{align}
which complement eq.~(\ref{513}) in the ghost sector. With these conditions, the renormalized 
integral equation system at zero temperature 
becomes\footnote{The finite parts in the ghost counter term $\delta Z_c$ are such that  
the first equation in (\ref{514}) could also be obtained by simply subtracting the bare 
ghost DSE for $\eta_0(\ka)$ at the renormalization scale $\mu_c$.} 
\begin{align}
\eta_0(\ka)^{-1} &= \eta_0(\mu_c)^{-1} - N g^2\,\Big[I_\eta^{(0)}(\ka) - I_\eta^{(0)}(\mu_c)\Big]
\label{514}\\[3mm] 
\bar{\omega}_0(\ka) &= Z\,\frac{\mu^2 - M_A^2}{\mu^2 - \mu_0^2}\,k^2+ Z\,\frac{M_A^2-\mu_0^2}
{\mu^2 - \mu_0^2}\,\mu^2 + 
\frac{N g^2}{\mu^2 - \mu_0^2}\,\Big[ \mu^2\big(I_\chi^{(0)}(\ka) - I_\chi^{(0)}(\mu_0)\big) - 
\nonumber \\[2mm]
&\qquad\qquad\qquad\qquad \qquad\qquad\quad
- k^2 \,\big(I_\chi^{(0)}(\mu) - I_\chi^{(0)}(\mu_0)\big) - \mu_0^2\,\big(I_\chi^{(0)}(\ka) - 
I_\chi^{(0)}(\mu)\big)\Big]\,.
\nonumber
\end{align}

Note that the difference of integrals in the square brackets is finite: For the first equation
(\ref{514}), this is clear because $I_\eta^{(0)}$ is logarithmically UV divergent by power counting.
In the second equation (\ref{514}), $I_\chi^{(0)}$ is quadratically divergent, but the subleading 
logarithmic divergence is also eliminated due to the clever combination of $I_\chi^{(0)}$ at three 
different scales. 

\medskip\noindent
The system eq.~(\ref{514}) is well suited for a numerical evaluation. It contains, 
besides the three renormalization constants $\{\eta_0(\mu_c), Z, M_A^2 \}$, also the 
coupling strength\footnote{In the present approximation scheme, $Ng^2$ is a finite 
adjustable parameter since there are no vertex corrections.} in the combination $Ng^2$. 
However, this parameter is \emph{redundant}. 
To see this, consider a solution $\bar{\omega}_0(k),\,\eta_0(k)$ of 
eq.~(\ref{514}) obtained with a set of parameters $\{\eta_0(\mu_c), Z, M_A^2, Ng^2\}$.  
From the form of the loop integrals in eq.~(\ref{514}) listed in appendix \ref{app:loop}, 
it is apparent that the rescaled parameters
\begin{align}
\{\,\eta_0(\mu_c),\, Z,\, M_A^2,\, Ng^2\,\} \to \{\,a\cdot\eta_0(\mu_c),\, b \cdot Z,\, M_A^2,\, b/a^2\cdot Ng^2\,\}
\label{622}
\end{align}
lead to the solution $a\,\eta_0(k)$ and $b\,\bar{\omega}_0(k)$, in which the propagators are 
scaled by arbitrary factors. Thus, any change in $Ng^2$ can be compensated by a corresponding rescaling 
of the propagators (and vice versa), including their field renormalisation constants $\eta_0(\mu_c)$ and $Z$.  
We can thus \emph{choose} to set $Ng^2 = 1$ in eq.~(\ref{514}), with the provision that the propagators 
(and their field renormalization constants) may later be multiplied by an overall constant when fitting 
to lattice data. The solution of eq.~(\ref{514}) therefore depends on the three renormalization 
constants $ \{ \eta_0(\mu_c), Z, M_A^2 \}$ only, where the first two determine the overall scale 
of the propagators, while $M_A^2$ determines the behaviour in the intermediate momentum region. 
Formally, $M_A^2$ can be chosen arbitrarily, but it is related through a gap equation to the gluon 
propagator and the bare Mass $M_0$. Since this gap equation involes $Ng^2$, different colour groups 
$SU(N)$ will require, in general, different value for $M_A^2$. For the case of $SU(2)$, we will 
give a numerical estimate for $M_A^2$ from high-precision lattice data in section \ref{sec:5a} 
below.

\subsection{Renormalization at finite temperatures}
\label{sec:4c}
Next we turn to the integral equation system at finite temperatures. In the previous section, 
the counter terms were constructed at zero temperature and it is important to use the 
\emph{same} counter terms also at non-zero temperatures in order to compare different
temperatures reliably. To do so, we merely have to add $( -\delta  Z_c)$ from eq.~(\ref{513}) 
to the right hand side of eq.~(\ref{480}), and likewise $(\delta Z_A\,k^2 + \delta M^2)$ 
from eq.~(\ref{530}) to the right hand side of eqs.~(\ref{420}) and (\ref{430}). 
The resulting system of equations can be written in two equivalent forms. 
This is best demonstrated at the ghost equation:
\begin{align}
\eta(\ka)^{-1} &= 1 - Ng^2\,I_\eta(\ka) - Ng^2\,\frac{\vek{k}^2}{k^2}\,\underbrace{\int_\beta \dd q
\,\frac{\eta(k-q)}{(k-q)^2}\,\Big[1 - (\unitthree{k}\cdot\unitthree{q})^2\big]\,\Big[\wb(\qa)^{-1} - 
\sb(\qa)^{-1}\Big]}_{\equiv L_\eta(\ka)} - \delta Z_c
\nonumber \\[2mm]
&= 1 - Ng^2\,I_\eta(\ka) - Ng^2\,\frac{\vek{k}^2}{k^2}\,L_\eta( k) - \Big[1-Ng^2\,I_\eta^{(0)}(\mu_c)
- \eta_0(\mu_c)^{-1}\Big]
\nonumber \\[2mm]
&= \eta_0(\mu_c)^{-1} - Ng^2\,\Big[ I_\eta( k) - I_\eta^{(0)}(\mu_c)\Big] - 
Ng^2\,\frac{\vek{k}^2}{k^2}\,L_\eta( k)
\nonumber \\[2mm]
% &=  \eta_0(\mu_c)^{-1} - Ng^2\,\Big[ I_\eta^{(0)}(\ka) - I_\eta^{(0)}(\mu_c)\Big]
% - Ng^2\,\Big[ I_\eta(\ka)-I_\eta^{(0)}(\ka)\Big] - Ng^2\,\frac{\vek{k}^2}{k^2}\,L_\eta(\ka)
% \nonumber \\[2mm]
&=  \eta_0(\ka)^{-1} - Ng^2\,\Big[ I_\eta(\ka)-I_\eta^{(0)}(\ks)\Big] - 
Ng^2\,\frac{\vek{k}^2}{k^2}\,L_\eta(\ka)\,,
\label{800}
\end{align}
where the zero temperature equation (\ref{514}) was used in the last step.
The advantage of this formulation is 
that it avoids all explicit renormalization factors, as the finite temperature solution is 
expressed as its zero temperature counterpart plus a temperature-dependent correction which 
must be finite  as all counterterms have already been consumed in $\eta_0(\ka)$. 
To confirm this, we note that the curvature and mass term in the gap equations (\ref{420}), (\ref{430}) 
are subdominant, which implies that $|\wb(\ka)^{-1} - \sb(\ka)^{-1}| \sim \mathcal{O}(k^{-4})$ at 
large momenta and $L_\eta(\ka)$ is finite by power counting. As for the difference
$I_\eta(k)-I_\eta^{(0)}(k)$ of logarithmically divergent loop integrals,
we note that this would be finite at \emph{zero temperature}. At nonzero temeprature, however,
this statement is less obvious, since $I_\eta(k)$ and $I_\eta^{(0)}(k)$ involve different 
form factors in their integrand. In principle, the restauration of $O(4)$ invariance 
at large momenta should ensure that the difference is finite at any temperature, but this 
limit is very hard to reach \emph{numerically}. We will therefore follow a different 
route in our numerical investigation below. 

Similar considerations can also be applied to the gap equation. After some algebra, we obtain
\begin{alignat}{3}
\eta(k_0,|\vek{k}|)^{-1} &= \eta _0(\ks)^{-1} & 
&- Ng^2\,\Big[ I_{\eta}(k_0,|\vek{k}|) - I_{\eta}^{(0)}(\ks)\Big]   & 
&- \frac{\vek{k}^2}{k_0^2 + \vek{k}^2} \,Ng^2\,L_{\eta}(k_0,|\vek{k}|)
\nonumber\\[2mm]
\wb(k_0,|\vek{k}|) &= \bar{\omega}_0(\ks) &
&+ Ng^2\,\Big [ I_\chi^\perp(k_0,|\vek{k}|) - I_\chi^{(0)}(\ks)\Big] &
&+ \Big[ M_\perp^2(\beta) -M_0^2 \Big]
\label{1000} \\[2mm]
\sb(k_0,|\vek{k}|) &=\bar{\omega}_0(\ks) &
&+ Ng^2\,\Big [ I_\chi^\|(k_0,|\vek{k}|) - I_\chi^{(0)}(\ks)\Big] & 
&+ \Big[ M_\perp^2(\beta) -M_0^2 \Big]+ \frac{\vek{k}^2}{k_0^2 + \vek{k}^2} \,\Ml^2(\beta)\,.
\nonumber 
\end{alignat}
To avoid confusion, we have reinserted to full momentum dependence in the 
finite-tem\-pe\-ra\-ture profiles and loop integrals.  Explicit expressions for the loop 
integrals entering this system can be found in appendix \ref{app:loop}.

\medskip
As mentioned above, the system eq.~(\ref{1000}) is hard to treat \emph{numerically} because the 
difference of loop integrals involve non-zero and zero temperature form factors, respectively. 
We will therefore rewrite these equations to bring them as close as possible to the 
zero-temperature case. Let us again consider the ghost equation as an example. 
From the second line in eq.~(\ref{800}), we have
\begin{align*}
\eta(k_0,|\mathbf{k}|)^{-1} &= \eta(0,\mu_c)^{-1} - 
Ng^2\,\Big[I_\eta(k_0,|\mathbf{k}|) - I_\eta(0,\mu_c)\Big] - 
Ng^2\,\Big[\frac{\mathbf{k}^2}{k_0^2 + \mathbf{k}^2}\,L_\eta(k_0,|\mathbf{k}) - 
L_\eta(0,\mu_c)\Big]
\end{align*}
Here, we have chosen the renormalization point at vanishing Matsubara 
frequency.\footnote{The orientation of the renormalization point in $3$-space is 
arbitrary due to the residual rotation symmetry, and we can choose the same 
direction as the external momentum, $(0,\mu_c\cdot \hat{\mathbf{k}})$, if this is 
ever necessary. A similar statement applies to the other two renormalization 
scales $\mu_0$ and $\mu$.} 
The benifit of this reformulation is that it only involves the difference
of \emph{the same} finite-temperature loop integrals (with the same form factors) 
at different scales. The cumbersome difference of loop integrals with different profiles
is now hidden in a single number, namely the (finite) value of the form factor at 
the renormalization scale, 
\begin{align}
\eta(0,\mu_c)^{-1} = \eta_0(\mu_c)^{-1} - \Big[ I_\eta(0,\mu_c) - I_\eta^{(0)}(\mu_c)
 - L_\eta(0,\mu_c) \Big]\,. 
\label{800b}
\end{align}
Again, we cannot compute this quantity reliably due to numerical issues, 
and we cannot choose it arbitrarily either, as it is determined completely by 
the zero-temperature counter terms. We will instead fix $\eta(0,\mu_c)$
indirectly as explained in detail in section \ref{sec:5b} below.  

The gap equations can be rewritten in much the same way. After some lengthy 
algebra, we arrive at the following system:
\begin{align}
\eta(k_0,k)^{-1} &= \eta(0,\mu_c)^{-1} - 
Ng^2\,\Big[I_\eta(k_0,k) - I_\eta(0,\mu_c)\Big] - 
Ng^2\,\Big[\frac{k^2}{k_0^2 + k^2}\,L_\eta(k_0,k) - L_\eta(0,\mu_c)\Big]
\nonumber\\[2mm]
\wb(k_0,k) &= 
\wb(0,\mu) + \frac{\mu^2 - (k_0^2 + k^2)}{\mu^2 - \mu_0^2}\,
\Big[\wb(0,\mu_0) - \wb(0,\mu) \Big] + \nonumber \\[2mm]
&\qquad + \frac{Ng^2}{\mu^2 - \mu_0^2}
\Bigg\{ \mu^2 \,\Big[ I_\chi^\perp(k_0,k) - I_\chi^\perp(0,\mu_0)\Big]
- (k_0^2 + k^2) \,\Big[ I_\chi^\perp(0,\mu) - I_\chi^\perp(0,\mu_0)\Big]
\nonumber \\[2mm]
&\hspace*{3cm} - \mu_0^2 \,\Big[ I_\chi^\perp(k_0,k) - 
I_\chi^\perp(0,\mu)\Big] \Bigg\}
\nonumber \\[2mm]
\sb(k_0,k) &= 
\sb(0,\mu) + \frac{\mu^2 - (k_0^2 + k^2)}{\mu^2 - \mu_0^2}\,
\Big[\sb(0,\mu_0) - \sb(0,\mu) \Big] - 
\frac{k_0^2}{k_0^2 + k^2}\,\Ml^2(\beta) +
\nonumber \\[2mm]
&\qquad + \frac{Ng^2}{\mu^2 - \mu_0^2}
\Bigg\{ \mu^2 \,\Big[ I_\chi^\|(k_0,k) - I_\|^\perp(0,\mu_0)\Big]
- (k_0^2 + k^2) \,\Big[ I_\chi^\|(0,\mu) - I_\chi^\|(0,\mu_0)\Big]
\nonumber \\[2mm]
&\hspace*{3cm} - \mu_0^2 \,\Big[ I_\chi^{\|}(k_0,k) - 
I_\chi^\|(0,\mu)\Big] \Bigg\}\,.
\label{800x}
\end{align}
Here, $k=|\mathbf{k}|$ and we have spelled out all arguments to the loop integrals
for clarity. The gap equations now exhibit the same combination of curvature integrals 
at three different scales which eliminated all divergences in the zero-temperature 
case. Alltogether, eq.~(\ref{800x}) contains, in addition to the coupling 
constant $Ng^2$, five temperature-dependent constants,
\begin{align}
\wb(0,\mu_0)\,,\qquad\quad \wb(0,\mu_0)\,,\qquad\quad\sb(0,\mu_0)\,,\qquad\quad 
\sb(0,\mu_0)\,,\qquad\quad\eta(0,\mu_c)\,,
\label{800y}
\end{align}
which are all determined, in principle, by the zero-temperature counter terms through 
finite equations that are, however, hard to treat numerically. For completeness, these 
relations are listed in appendix \ref{app:constants}, although we will not use them and 
determine the constants in eq.~(\ref{800y}) indirectly by other means.

The systems (\ref{1000}) and (\ref{800x}) are the main result of this paper. In the next chapter, 
we will give details on its numerical implementation and discuss the solutions.

\section{Numerical treatment and Results}
\label{sec:5}
\subsection{Renormalization at zero temperature}
\label{sec:5a}
As explained in the last section, the renormalization of our approach must be carried 
out entirely at zero temperature. In section \ref{sec:4b}, we have presented a 
simplified procedure, where all counter terms are determined by simple local 
prescriptions to low order Green's funcitons. Since this is different from the 
procedure initially layed out in Ref.~\cite{Quandt:2013wna}, we cannot directly 
take over those results and must first re-determine the optimal renormalization 
parameters to match the known lattice results. This is carried out at 
zero temperature, and we take again the high-precision $SU(2)$ lattice data from 
Ref.~\cite{Bogolubsky:2009dc}.

\begin{figure}
\begin{center}
\includegraphics[width=11cm]{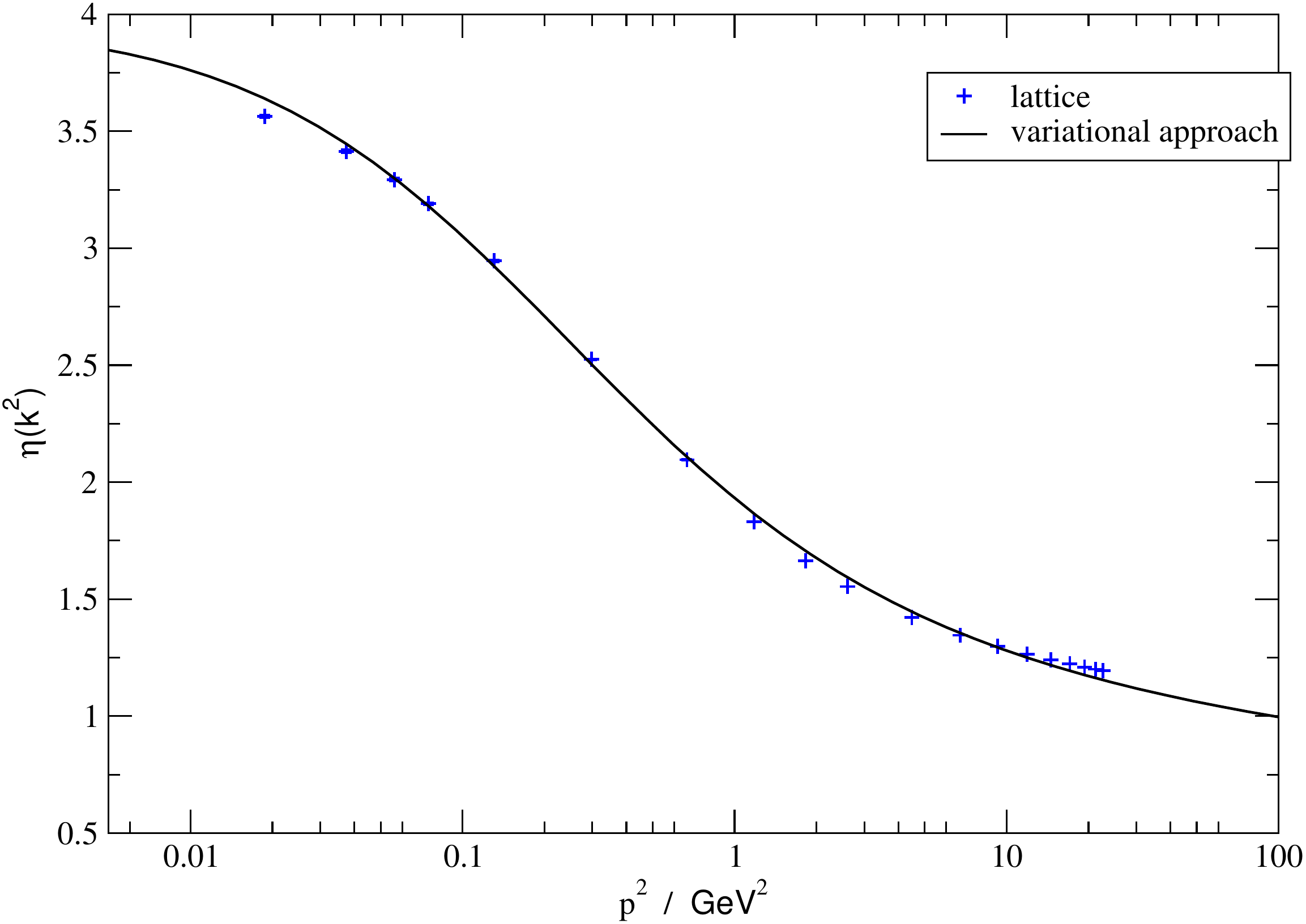}
\\[6mm]
\includegraphics[width=11cm]{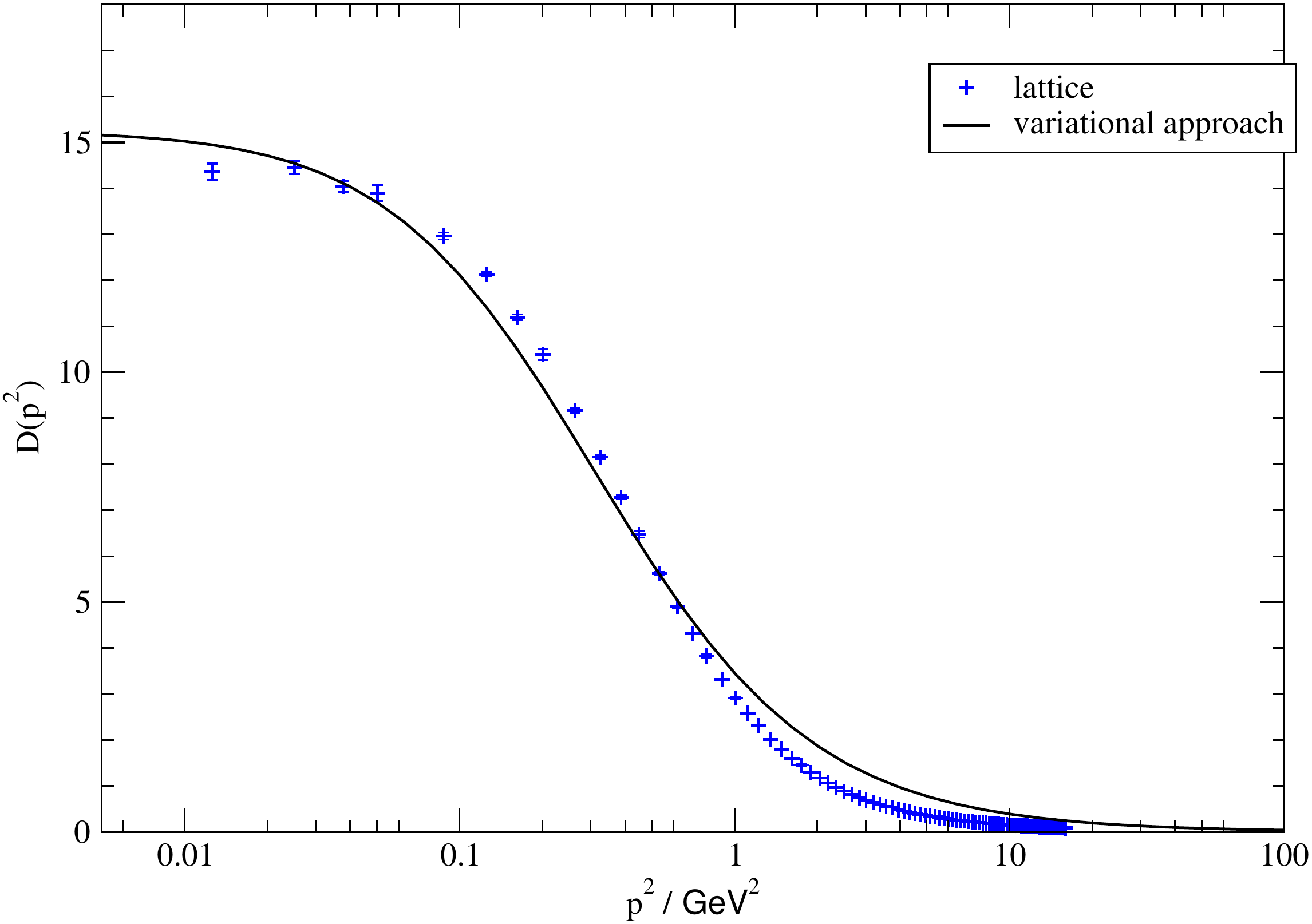}
\end{center}
\caption{The ghost form factor $\eta(p^2)$ (\emph{top}) and the gluon 
propagator $D(p^2)$ (\emph{bottom}) in Landau gauge at zero temperature.
Lattice data is taken from Ref.~\cite{Bogolubsky:2009dc}, and the
gluon mass renoramalization parameter is $M_A \approx 0.48\,\text{GeV}$.}
\label{fig:1} 
\end{figure}

In figure \ref{fig:1}, we compare the best fit of our zero-temperature formulation to the 
lattice data for both the gluon propagator and the ghost form factor. Since we already 
know that the physically realized solution is \emph{subcritical} (decoupling solution), 
only this type of solution is investigated. The scale $\mu_0$ for the gluon mass 
renormalization could thus be taken to zero, but it is numerically favourable 
to impose this  condition at a small but non-zero scale  $\mu_0 \approx \mu / 10$. 

As can be seen, the new renormalization prescription describes the lattice data with 
similar (high) precision as the more complicated prescription in 
Ref.~\cite{Quandt:2013wna}. We find $M_A / \mu  \approx 0.49$ at $\mu \approx 1\,\text{GeV}$
to be a favourable parameter, but this determination has rather great uncertainties. 
As can be seen from figure~\ref{fig:1}, this is not because the approach could not 
describe the data, but rather since different combinations of the renormalization 
constants produce similar results, i.e.~there are (almost) flat directions in 
renormalization space. From the comparision to the lattice data, we can also 
determine our parameters in absolute numbers. For the choice in figure \ref{fig:1}, 
we can match the data at the point $\mu =0.96 \,\text{GeV}$, where we 
find 
\begin{align}
M_A \approx 0.48\,\mathrm{GeV}\,. 
\label{a100}
\end{align}
The renormalization scale $\mu$ is, of course, arbitrary as long as it is sufficently 
deep in the ultra-violet. 
To check this, we have re-determined the renormalization parameters 
when matching at the larger scale $\mu = 5 \,\text{GeV}$. The best match to the lattice 
data in this case is virtually undistinguishable from figure \ref{fig:1}. The 
renormalization parameters change considerably and, in particular, we now have 
$M_A / \mu \approx 0.173$, which again yields eq.~(\ref{a100}) in physical units. 
In this sense, the mass parameter $M_A$ is invariant under the renormalization group.

It must be emphasized, however,  that the determination of the absolute scales 
is not very accurate  and different values may produce similar results as in 
figure~\ref{fig:1}. Ultimately, the scales should be fixed by the comparision 
to the string tension computed from the effective action for the \emph{Polyakov loop}, 
which is currently under investigation.

\subsection{Numerical procedure at finite temperature}
\label{sec:5b}
At finite temperature, our numerical procedure is based on eq.~(\ref{800x}). The five
constants eq.~(\ref{800y}) entering this system are, in principle, fixed through the 
zero-temperature counter terms $\eta_0(\mu_c)$, $Z$ and $M_A$. The first two 
of these three parameters only adjust the overall normalization of the 
zero-temperature propagators, which in turn fix the normalization of 
the finite-temperature propagators as well. 
In practice, the field normalization is rarely expressed through the 
zero-temperature counter terms, but rather by the alternative condition 
that the propagators at \emph{all} temperatures should pass through the same point 
at a very large scale\,, 
\begin{align}
\wb(0,\mu_\infty) = \sb(0,\mu_\infty) = Z\,\mu_\infty^2 \,,
\qquad\qquad \mu_\infty \gg 1\,.
\label{kain}
\end{align}
This is based on the idea that finite temperature effects should become
immaterial in the deep perturbative region where $O(4)$ invariance is restored. 
The lattice study in Ref.~\cite{Aouane:2011fv}, to which we compare, employs a similar 
prescription with $\mu_\infty = 5 \,\rm{GeV}$. If we adopt eq.~(\ref{a100}), 
this means $\mu_\infty \approx 5 \mu$, but much larger values are also accessible 
in our case. In any case, the prescription eq.~(\ref{kain}) can be imposed simply 
by replacing $\mu \to \mu_\infty$ in the second and third eq.~(\ref{800x}). 
The values at the reference scale, $\wb(0,\mu)$ and $\wb(0,\mu)$ are then 
temperature-dependent \emph{predicitons} of the calculation, which should agree
with the results of the more complicated approach \ref{app:constants} if the 
scale $\mu_\infty$ is chosen large enough. This reasoning allows us to compare 
to lattice data conveniently, and to determine the two constants 
$\wb(0,\mu)$ and $\wb(0,\mu)$ in eq.~(\ref{800x}) self-consistently.

For the ghost form factor, we would like to use the same procedure but this is 
hampered by the fact that we \emph{must} impose a boundary condition in the 
deep infra-red in order to discern scaling and decoupling types of solution.
This means that we have to \emph{guess} the (temperature-dependent)
intercept $\eta(0,\mu_c)$ at $\mu_c \approx 0$ such that the solution passes 
through the common value
\begin{align}
\eta(0,\mu_\infty) = Z_c
\label{abel}
\end{align}
for all temperatures. 
% In practice, we can make this guess a bit more educated:
% we start with an arbitrary value for $\eta(0,\mu_c)$ at each temperature and 
% compute $\eta(0,\mu_\infty)$. This will, in general, differ from eq.~(\ref{abel}) 
% and we re-adjust the initial condition $\eta(0,\mu_c) \to \eta(0,\mu_c) 
% \cdot Z_c / \eta(0,\mu_\infty)$ and recompute the solution. After a few 
% iterations, the correct intercept $\eta(0,\mu_c)$ will be found.
This leaves us with the remaining mass terms
\begin{align}
\wb(0,\mu_0) = Z\,m_\perp^2\,,\qquad\qquad
\sb(0,\mu_0) = Z\,m_{\|}^2\,,\qquad\quad
\Ml^2 = Z\,\Delta m_{\|}^2\,.
\label{eva}
\end{align}
which cannot be determined self-consistently by any scaling procedure, as 
they are not related to field renormalizations.
%\footnote{The  zero-temperature limit is $m_\perp^2 \to M_A^2$, $m_{\|}^2 \to M_A^2$ 
% and $\Delta m_{\|}^2 \to 0$.}
We will determine them  by fitting to the lattice data, i.e.~we have three 
parameters to cover the entire momentum range for all three propagators at each 
temperature. The need to fit eq.~(\ref{eva}) (when we actually have equations 
to compute them) is only due to numerical issues. We can, however, take the 
alternative point of view that the coefficients eq.~(\ref{eva}) are part of 
our variational ansatz (which includes the renormalization procedure), and 
the fit corresponds to a solution of the corresponding gap-equation. 
This is very much in line with the general philosophy of our approach, although 
it would, of course, be desirable to compute eq.~(\ref{eva}) self-consistently 
without input from the lattice.

It should also be mentioned that the mass parameter $\Ml^2$ could actually be 
computed from eq.~(\ref{1001}). We have still included it in the 
set of fit parameters, because eq.~(\ref{1001}) as well as the full dynamical 
system eq.~(\ref{800x}) contains the coupling constant $Ng^2$. From eq.~(\ref{800x})
and the explicit form of the loop integrals, it is apparent that the factors of 
$Ng^2$ could be eliminated by the same scaling procedure that we used in the case of 
zero temperature. However, $Ng^2$ is temperature-independent, and so, only 
temperature-independent norm changes (rescalings) of the propagators are allowed. 
In addition, such rescalings do not leave the mass coefficients in eq.~(\ref{eva}) 
invariant. This means that all three mass coefficients must be \emph{fitted} to 
lattice data  (or ideally determined from appendix \ref{app:constants}) if we 
rescale the propagators or change $Ng^2$. In particular, the mass coefficients 
eq.~(\ref{eva}) must be re-determined if we change the colour group $SU(N)$. 
The bottom line is that we are entitled to \textbf{(i)} set $Ng^2 = 1$,
\textbf{(ii)} re-determine the coefficients in eq.~(\ref{eva}) for each temperature 
and \textbf{(iii)} allow for temperature-independent rescalings of the 
propagators.
% \footnote{The norm for $\wb$ and $\wb$ must be scaled by the same factor 
% to ensure that the rescaled profiles are still a solution.} 
Since the mass parameter $\Delta m_{\|}$ has negligable influence on the final result, 
we will always set $\Delta m_{\|}=0$.
% \footnote{The suppression of the term with $\Ml^2$ is a consequence 
% of the prefactor $k_0^2 / (k_0^2 + k^2)$ which multiplies it: For large $k$, 
% this is suppressed against the leading behaviour $\sb(k_0,k) \sim (k_0^2 + k^2)$ 
% while the deep infrared is dominated by the lowest Matsubara frequency $k_0 = 0$, 
% to which the term does not contribute.}
Then, the entire system has only two free parameters, $m_\perp$ and $m_\|$, at each 
temperature, in addition to a temperature-independent overall norm scale for 
each form factor. 

\bigskip\noindent
Numerically, the main issue in eq.~(\ref{800x}) as compared to the zero-temperature case 
is the appearance of the Matsubara sum which replaces the frequency integral. There 
are various techniques to compute this sum. A direct evaluation is hampered by the 
fact that only a relatively small number of frequencies can be included in the coupled 
integral equation system to keep the overall computational effort under control. 
Typically, we include up to $n_{\rm max} \approx 20$ frequencies which corresponds,
at a typical temperature $\mu\beta \approx 1$, to a frequency cutoff of 
$2 \pi n_{\rm max} / \beta \approx 130 \mu$ which is much smaller than the 
spatial momentum cutoff $\Lambda/\mu = O(10^5)$.  Whether or not this is 
sufficient depends on the temperature itself: At high temperatures, the Matsubara sum 
converges quickly and summing $n_{\rm max}$ frequencies is sufficient. At lower 
temperatures, more frequencies contribute and we have to restort to more elaborate 
summation techniques as explained below.
% One method for intermediate relies on the observation that $O(4)$-invariance is 
% quickly restored at high momenta. We can therefore include a much larger number 
% of frequencies in the sum by \emph{extrapolating} the value of the higher 
% frequencies from the lower ones,
% \begin{align}
%  f(\nu_n,q) \to 
%  f(\nu_{\rm max},\sqrt{q^2 + \nu_n^2 - \nu_{\rm max}^2})\,,
%  \qquad \qquad n > n_{\rm max}\,,
%  \label{a101}
% \end{align}
% where $f$ stands for any of the profiles $\{\eta,\wb,\sb\}$. 
% The value of the loop integral after integrating over 
% angles. 
% 
% It should be emphasized again that the iterative determination of the 
% renormalization constants is only needed if their temperature-dependence 
% matters, i.e.~if we are \emph{comparing} different temperatures. 
% For general studies at a fixed temperature, we can just view them 
% as input parameter that can be set to any finite value. 

\subsection{Cutoff independence}
\label{sec:5c}
Before presenting our numerical results, we must ensure that the finite-temperature 
system eq.~(\ref{800x}) is indeed cutoff-independent. We must show this for both 
the momentum and frequency cutoff independently, and for all components of the 
solution at various temperatures. This gives a bewildering number of plots and 
we only present two representative cases, the remaining ones all displaying the 
same behaviour. 

Since the cutoff-dependence can be studied at each temperature 
independently, we do not need to guess the correct ghost intercept
$\eta(0,\mu_c)$ to ensure eq.~(\ref{abel}) at all temperatures. Instead, 
we take an arbitrary fixed value $\eta(0,\mu_c) = 14$ at $\mu_c = 0$. 
The gluon normalization is fixed at $\mu_\infty = 4\,\mu$ with $Z=0.8$
and for the mass parameters, we take $m_\perp = m_{\|} = M_A \approx 0.5\,\mu$ 
at $\mu_0 = \mu / 10$ (and $\Delta m_{\|} = 0$).
With this setup, we study the ghost form factor $\eta(k_0,|\mathbf{k}|)$ as a 
function of the spatial momentum  $k = | \mathbf{k}|$, at vanishing Matsubara 
frequency $k_0 = 0$.\footnote{Higher Matsubara frequencies $k_0 > 0$ show 
the same behaviour.} We vary both the frequency and momentum cutoff, at
various temperatures, and check how this affects the results. 

\renewcommand{\arraystretch}{1.2} 
\begin{table}[t!]
\centering
\begin{tabular}{c|ccccccc}
\toprule
 & $\,\,k=0.01\,\,$ & $\,\,k=0.05\,\,$ & $\,\,k=0.1\,\,$ & $\,\,k=0.5\,\,$&  
$\,\,k=1.0\,\,$ & $\,\,k=5.0\,\,$ & $\,\,k=10.0\,\,$ \\ \colrule
$\Lambda=1$     & 11.03  & 6.30 & 4.41 & 1.88 & 1.51 &      &       \\
$\Lambda=10$    & 11.03  & 6.30 & 4.41 & 1.88 & 1.50 & 1.25 &  1.22 \\
$\Lambda=100$   & 11.03  & 6.30 & 4.41 & 1.87 & 1.50 & 1.24 &  1.21 \\
$\Lambda=10000$ & 11.01  & 6.28 & 4.41 & 1.88 & 1.51 & 2.25 &  1.22
\\ \botrule
$\Lambda=1$     & 13.82  & 13.09 & 12.33 & 9.09  & 7.69  &      &       \\
$\Lambda=10$    & 13.82  & 13.09 & 12.33 & 9.12  & 7.65  & 5.53 &  5.00 \\
$\Lambda=100$   & 13.82  & 13.09 & 12.34 & 9.21  & 7.82  & 5.72 &  5.19 \\
$\Lambda=10000$ & 13.82  & 13.09 & 12.34 & 9.20  & 7.78  & 5.68 &  5.20
\\ \botrule
\end{tabular}
\caption{\label{tab1}Dependence of the ghost form factor $\eta(0,k)$ on the spatial 
momentum cutoff $\Lambda$. The upper table is for $\beta = 0.5$, the lower one 
for $\beta = 20.0$. All dimensionfull quantities are measured in units of the 
renormalization scale $\mu$.}
\end{table}
\renewcommand{\arraystretch}{1.0}

In table \ref{tab1}, we show the dependence of the ghost form factor on the 
spatial cutoff $\Lambda$ for two typical temperatures $\beta = 0.5 / \mu$ 
and $\beta = 20.0 / \mu$. We have fixed the number of Matsubara frequencies 
to $n_{\rm max} = 20$ and perform the frequency sum naively without further 
extrapolation. (We will see below that this is sufficient for these temperatures.)
As can be clearly seen, the dependence on the spatial momentum cutoff $\Lambda$ 
is almost negligable. The sensitivity is slightly more pronounced for the lower 
of the two temperatures, and for larger external momenta $k = | \mathbf{k} |$. 
In both cases, the maximal change of the ghost form factor $\eta(0,k)$ observed 
at $k=10 \mu$ is less than $4\%$ when varying $\Lambda$ by three orders of magnitude. 
This clearly indicates that both the quadratic \emph{and} logarithmic divergences 
have been removed in our formulation eq.~(\ref{800x}), and that the perturbative 
region sets in fairly quickly, even below $\Lambda = 10 \mu$.

\begin{figure}
\includegraphics[width=7cm]{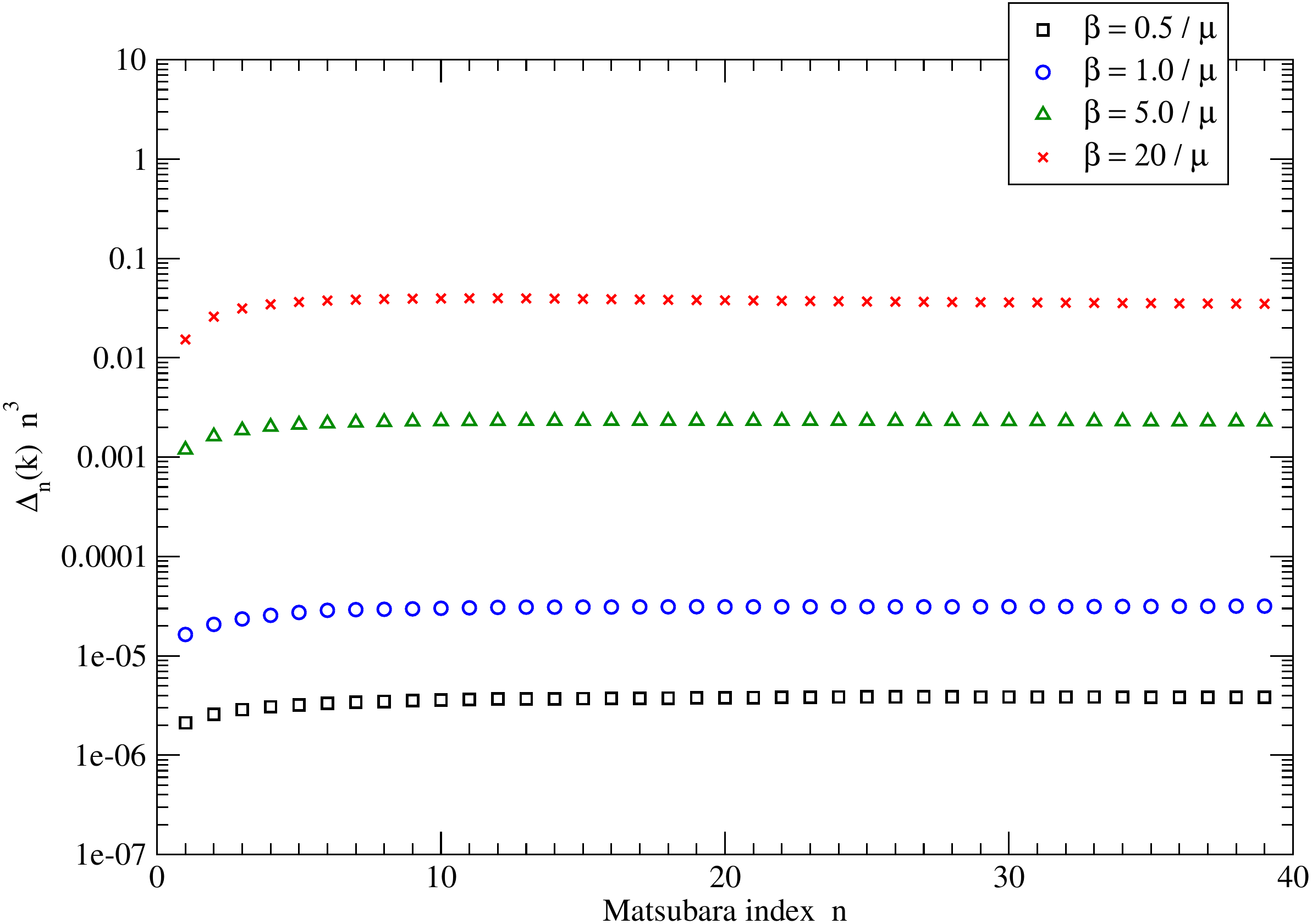}
\hspace*{1cm}
\includegraphics[width=7cm]{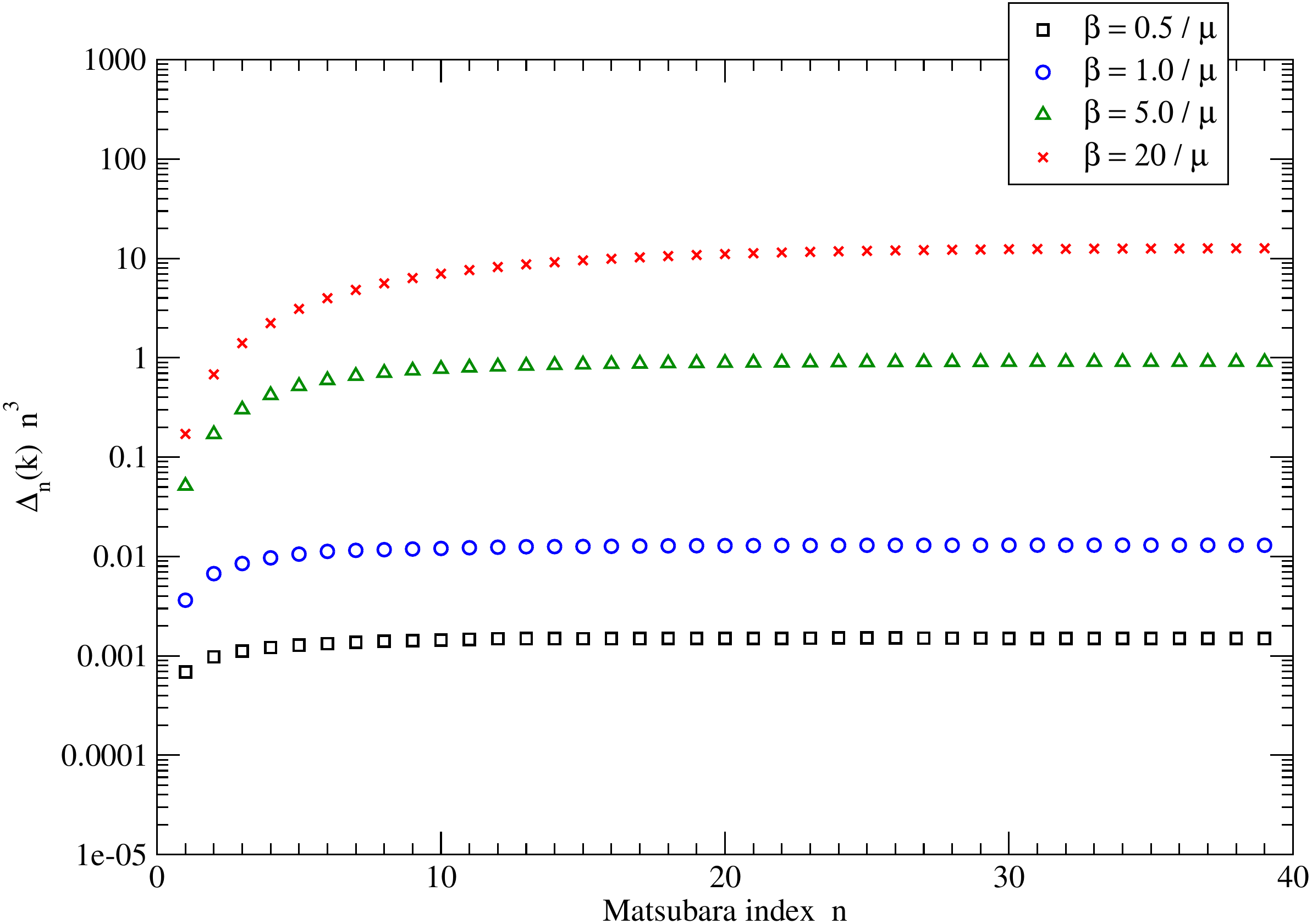}
\caption{The contribution $\Delta_n(k)$ of the $n$'th Matsubara mode to the 
gluon form factor $\eta(0,k)$ at external momentum $k$. The left panel shows 
data for $k=0.5\,\mu$ and various temperatures on a logarithmic plot, while the 
right panel shows the same for $k = 10\,\mu$. The contributions are 
multiplied by $n^3$ to better show their asymptotics.}
\label{fig:2} 
\end{figure}

\medskip
For the Matsubara sum, the situation is slightly more complicated. 
In general, the loop contributions from the Matsubara frequency \#$n$, after integrating 
over angles and spatial loop momentum, decay as $1/n^3$ asymptotically, which leads to a 
convergent Matsubara sum.\footnote{Our counting is such that $n \ge 0$ and contributions 
with $n \neq 0$ are actually the sum of $n$ and $(-n)$, which are equal in the 
present case.} To demonstrate this behaviour, let us consider the loop contribution
$\Delta_n(k)$ to the ghost form factor from the first line in eq.~(\ref{800x}),
\begin{align}
\eta(0,k)^{-1} - \eta(0,\mu_c)^{-1} &= \Big[I_\eta(0,\mu_c)- I_\eta(0,k)\Big] - 
\Big[L_\eta(0,\mu_c) - L_\eta(0,k)\Big]
\nonumber \\[2mm]
&\equiv \beta^{-1} \sum_{n=0}^\infty \int_0^\infty dq\,\int_{-1}^1 dz\, \cdots 
\equiv \sum_{n=0}^\infty \Delta_n(k)\,,
\end{align}
where the dots indicate the integrand of the corresponding loop integrals
(cf. appendix \ref{app:loop}), and we have put the external Matsubara 
frequency $k_0=0$ for convenience. In figure \ref{fig:2}, we plot $\Delta_n$ 
\emph{multiplied by} $n^3$, as a function of the Matsubara summation index $n$. 
As can be clearly seen, the asymptotic behaviour $\Delta_n(k) \sim n^{-3}$ is reached 
very quickly at small momenta (left panel, $k= 0.5\,\mu$), for all temperatures 
down to $\beta = 20 / \mu$. The same limit requires much more frequencies (up to $40$) 
at large external momenta  as demonstrated in the right panel ($k=10\,\mu$). 
The asymptotics of the individual contribution allows for an accurate estimate of 
the remainder in the infinite Matsubara series. However, the $n=0$ term still dominates 
in all cases,\footnote{This is not visible in the plot due to the factor $n^3$.} 
and the partial sum is found to always saturate once the asymptotic behaviour 
has set in. For instance, the worst convergence in figure \ref{fig:2} is at $k=10\,\mu$ 
for the lowest temperature $\beta = 20 / \mu$ (right panel). In this case, the asymptotics 
is not completely reached at $n=20$, but the partial sum of the first $20$ frequencies 
still agrees with the value extrapolated from the full asymptotic behaviour to better 
than $0.1 \%$. 

\begin{figure}
\includegraphics[width=7cm]{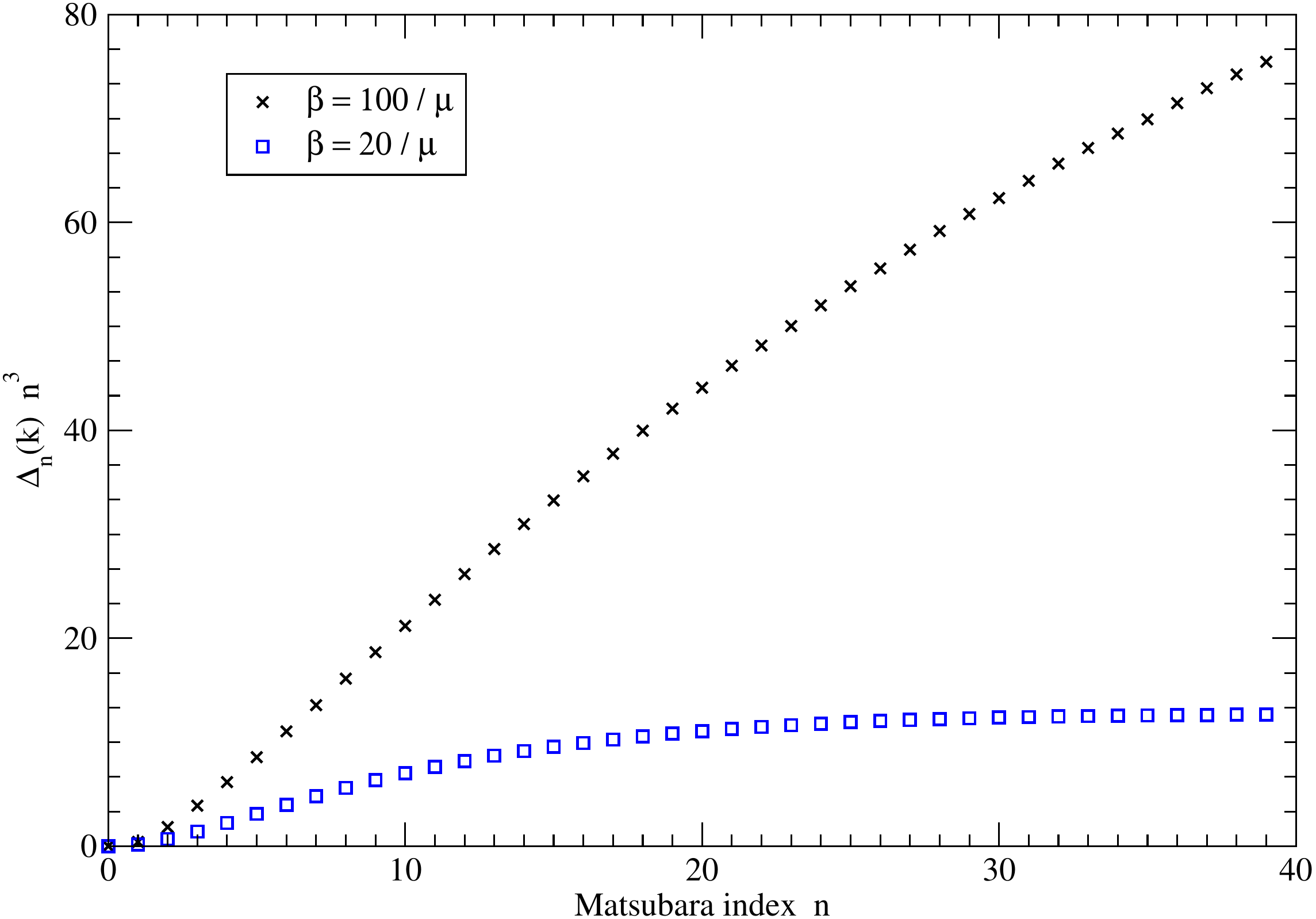}
\hspace*{1cm}
\includegraphics[width=7cm]{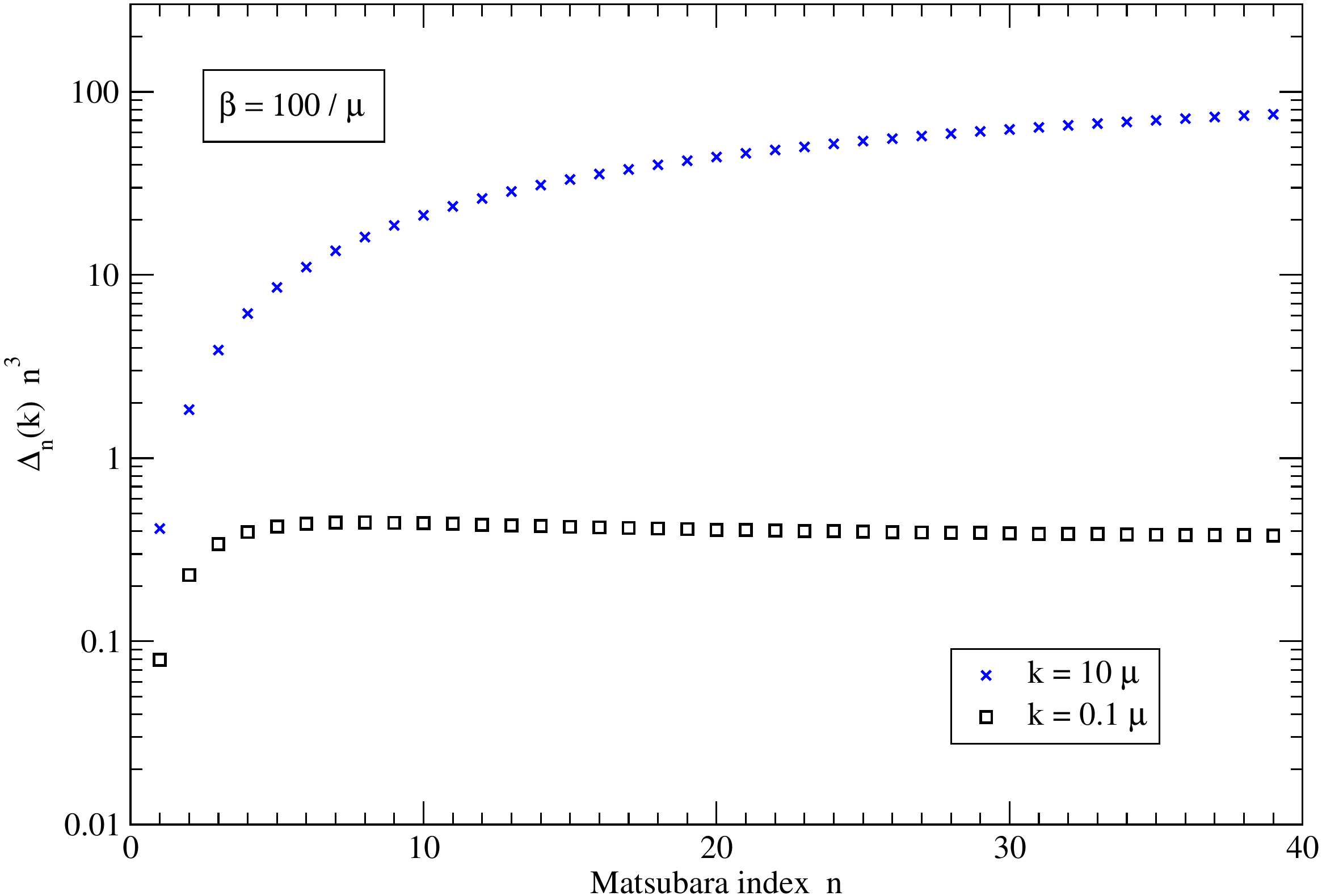}
\caption{The contribution $\Delta_n(k)$ of the $n$'th Matsubara mode to the 
gluon form factor $\eta(0,k)$ at external momentum $k$. The left panel shows 
data for a large momentum $k=10 \mu$ and two different temperatures on a linear plot, 
while the right panel displays data for a very low temperature $\beta = 100/ \mu$ 
at two different momenta $k$. Again, the contributions are multiplied by $n^3$ 
to better display their asymptotics.} 
\label{fig:3} 
\end{figure}

In figure \ref{fig:3}, we show again $\Delta_n(k)\,n^3$, but this time on a linear 
plot which exhibits the slower convergence to the asymptotics more clearly. In the left 
panel of figure \ref{fig:3}, we take a large momentum $k=10\,\mu$ and study 
two different temperatures. It is apparent that $n=30$ frequencies are enough to see the 
asymptotic behaviour and saturate the Matsubara sum at the intermediate temperature 
$\beta = 20 / \mu$. As we further lower the temperature to $\beta = 100 / \mu$, no 
sign of the correct asymptotics can be seen within the first $40$ frequencies, 
although the $n=0$ term still dominates in this case. The right panel of figure 
\ref{fig:3} shows that the convergence speed depends very much on the external 
momentum and, quite generally, decreases as the momentum is increased. 

\medskip
The bottom line is that summing up to $n=30$ frequencies directly is sufficient to 
saturate the Matsubara sum for all relevant momenta down to temperatures of about
$\beta = 20 / \mu$, though as few as $4$ frequencies are necessary at higher temperatures.
By contrast, very low temperatures can only be trusted for small momenta $k \ll \mu$,
and would require an excessive amount of CPU time otherwise.\footnote{Series accelerators 
also do not help in this case, since the eventual asymptotics has not been reached, 
cf.~fig.~\ref{fig:3}.} We will mainly restrict our numerical effort to include up to 
$n=20,\ldots,40$ frequencies at maximum, which means that we must content ourselves with 
temperatures not much below $\beta = 20 / \mu$.

\subsection{Comparision with lattice results}
\label{sec:5d}
\begin{figure}

\includegraphics[width=10cm]{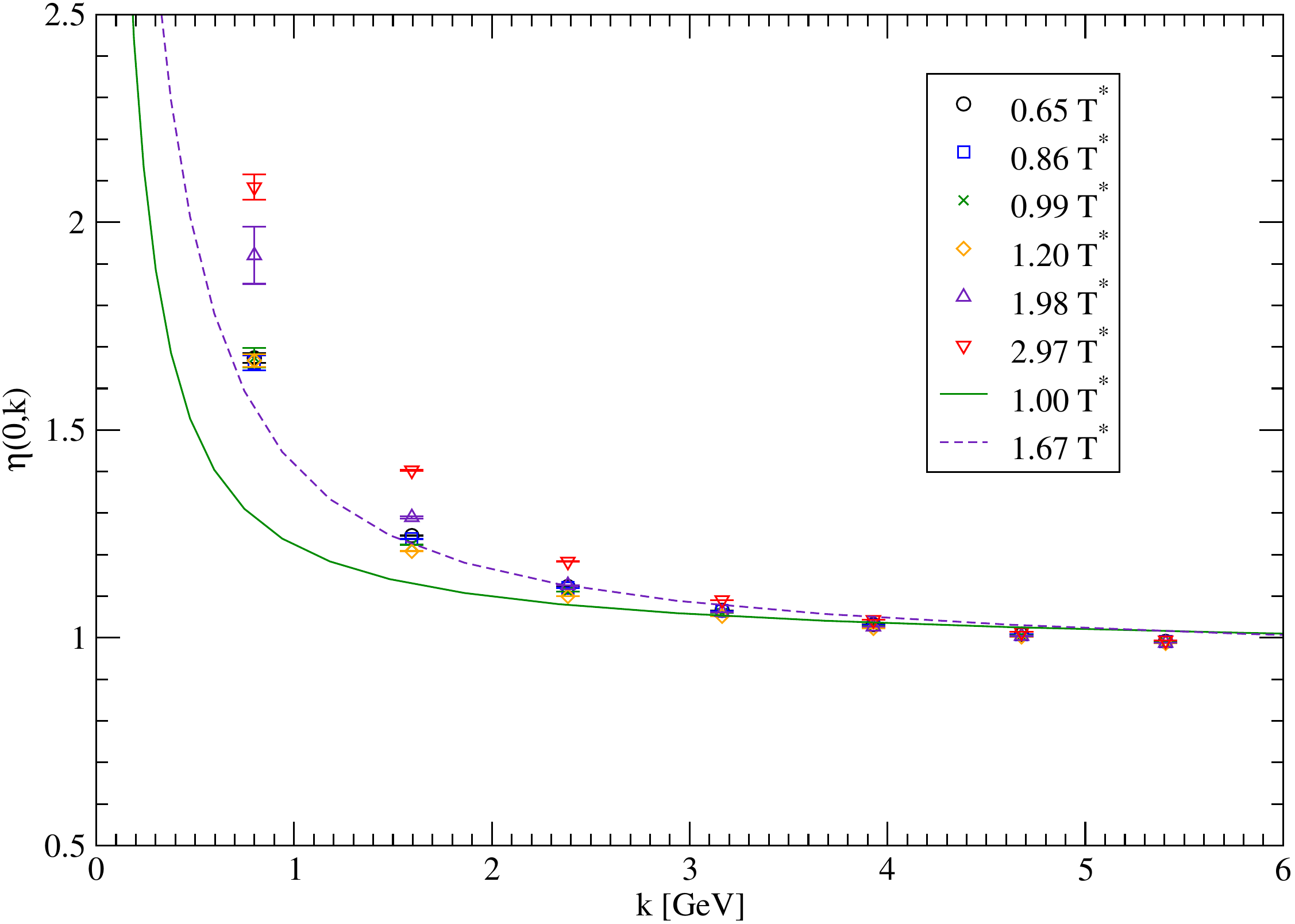}
\caption{
The renormalized ghost form factor at various temperatures, as a function
of the spatial momentum $k =|\mathbf{k} |$ with frequency $k_0 = 0$. 
Lattice data is taken from Ref.~\cite{Bogolubsky:2009dc}.
}
\label{fig:4} 
\end{figure}

\begin{figure}
\includegraphics[width=7cm]{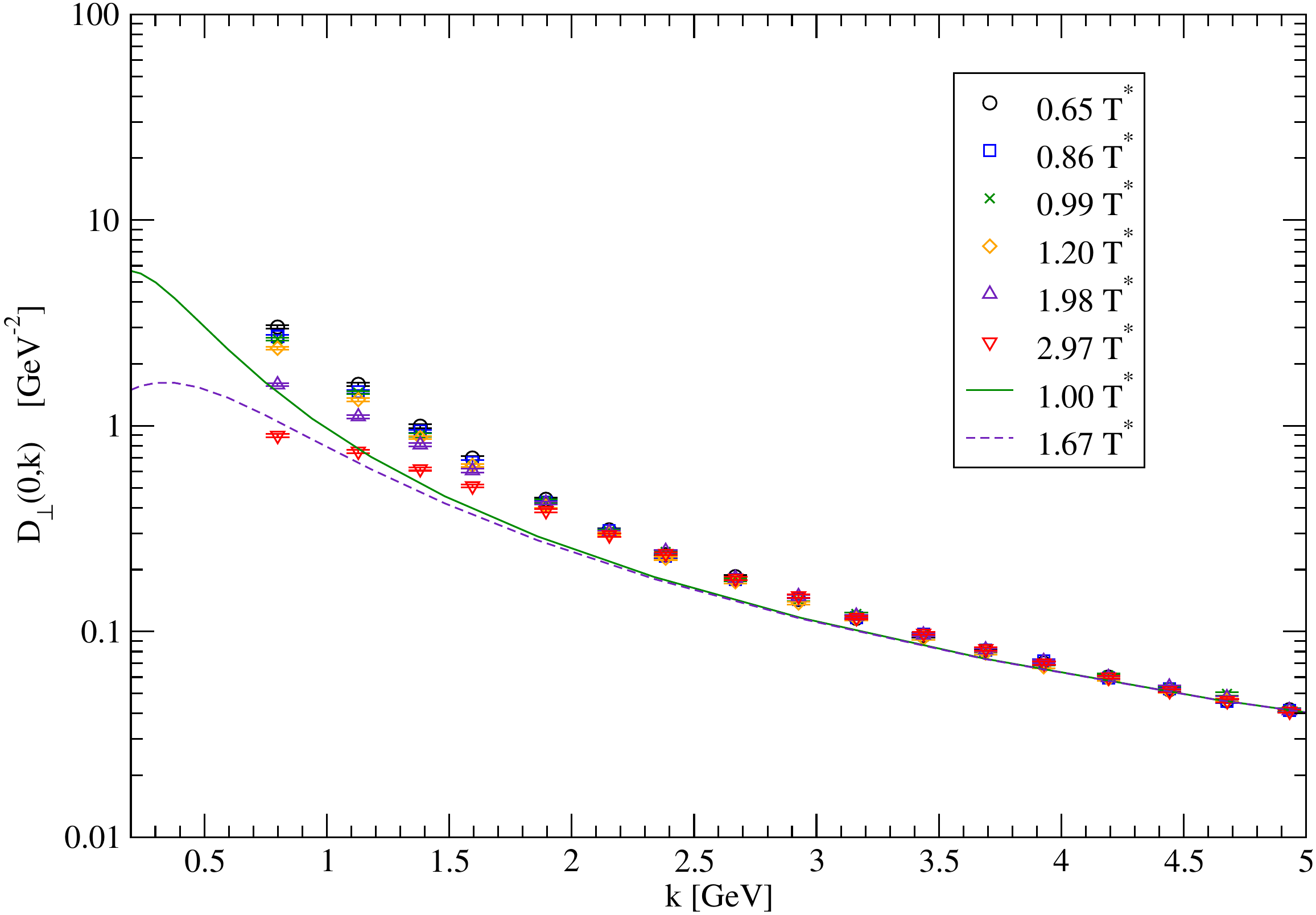}
\hspace*{1cm}
\includegraphics[width=7cm]{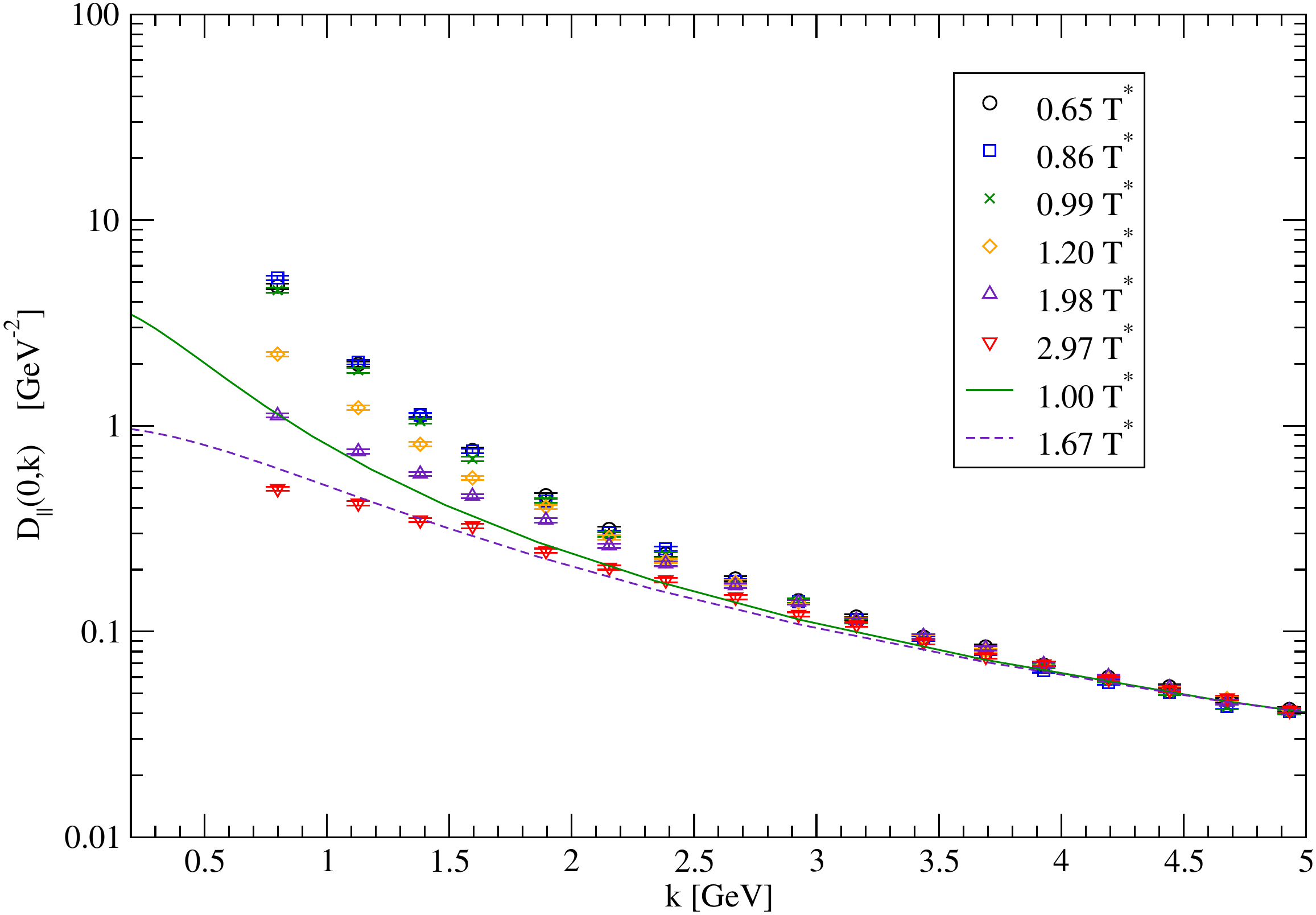}
\caption{The renormalized gluon propagator at various temperatures, as a function
of the spatial momentum $k =|\mathbf{k} |$ with frequency $k_0 = 0$. 
The left panel shows the transversal component $D_\perp(0,k) = \wb(0,k)^{-1}$, 
while the right panel shows the longitudinal component $D_{\|}(0,k) = \sb(0,k)^{-1}$. 
Lattice data is taken from Ref.~\cite{Bogolubsky:2009dc}.}
\label{fig:5} 
\end{figure}

Finally, figure \ref{fig:4} and \ref{fig:5} show our cummulative results 
for the ghost and gluon propagator, respectively, at various temperatures 
compared to the $SU(3)$ lattice data taken from Ref.~\cite{Bogolubsky:2009dc}. 
The lattice data was renormalized at a large scale $\mu_\infty = 5\,\text{GeV}$ 
in the perturbative region, with 
$\eta(0,\mu_\infty) = \mu_\infty^2\,D_\perp(0,\mu_\infty) 
= \mu_\infty^2\,D_{\|}(0,\mu_\infty) = 1$.

To translate this setting in our notation, we first set the arbitrary 
renormalization scale $\mu = \mu_\infty = 5 \,\text{GeV}$, because $\mu$ is 
anyhow redundant once we change our renormalization prescription to match the 
lattice convention. In these units, the critical temperature is at 
$\beta^\ast = 16.67$. The lattice data goes down to temperatures of about
$T = 0.67\,T^\ast$, which translates into $\beta = 25$ which, according to the 
studies in the previous section, is still treatable with reasonable numerical 
effort. The gluon propagator normalization becomes $Z = \omega(0,\mu_\infty) / \mu^2 
= 1 / ( D(0,\mu_\infty)\,\mu_\infty^2) = 1$. Finally, the intercept $\eta(0,\mu_c)$ 
is chosen at each temperature such that $\eta(0,\mu_\infty) = 1$ as discussed 
in the previous section. This fixes all the normalization, and leaves us 
with just the two fit parameters $m_\perp(\beta) / \mu$ and $m_{\|}(\beta) / \mu$ 
at each temperature.\footnote{We set $\Delta m_{\|} = 0$ throughout as it has no 
effect on the final results.}

As can be seen from the plots, our approach is able to reproduce the basic 
properties of the propagators in full Yang-Mills theory at finite temperatures:
\begin{enumerate}
 \item the gluon propator is suppressed as the temperature increases
 \item the ghost form factor is enhanced as the temperature increases
 \item the temperature sensitivity of the \emph{longitudinal} gluon propagator
 $D_{\|}(0,k)$ is much greater than for the transversal propagator
 (note the logarithmic scale in figure \ref{fig:5}).
\end{enumerate}

Quantitatively, the gluon propagator seems to miss some strength in the
intermediate to low momentum regime, while the ghost form factor is too steep
in this region. These problems seem, however, to be related to the determination 
of the overall scale $\mu = \mu_\infty$, which was taken from the match 
of the gluon propagator with the lattice data at $\mu = 5 \,\text{GeV}$. As can 
be seen from figure \ref{fig:5}, this determination is not very accurate as
the propagators lie on top of each other for all temperatures at least down to 
about $3\,\text{GeV}$. Thus, there is some uncertainty of almost a factor of 
$2$ in $\mu$, which translates in the other dimensionfull quantities such as 
the norm of the gluon propagator or the mass parameters. Eventually, this scale 
should better be fixed intrinsically by computing a dimensionfull quantity such 
as the string tension from the Polyakov loop potential.

%%%%%%%%%%%%%%%%%%%%%%%%%%%%%%%%%%%%%%%%%%%%%%%%%%%%%%%%%%%%%%%%%%%%%%%%%%%%%%%%%%%%%%%

\section{Summary and conclusions}
\label{sec:6}
In this paper, we have extended the covariant variational principle for Yang-Mills 
theory in Landau gauge to the case of non-zero temperatures. In the course of the 
derivation, we have also clarified and simplified the renormalization procedure 
at \emph{zero} temperature, and re-determined the gluon mass parameter $M_A$ 
(which is the only free parameter in our approach besides the overall normalization 
of the propagators) by comparision to lattice data. 
At finite temperatures, no further counter terms are necessary to put the system 
in a form that is manifestly independent of the UV cutoff, both in spatial 
and temporal momentum direction. We have carefully corroborated this cutoff-independence
through numerical studies. Two mass parameters, which in principle are fixed from the 
$T=0$ sector, had to be fitted to lattice data for practical reasons. 
The results are in qualitative agreement with recent high precision lattice data.
We have also discussed possible causes of the discrepancy, which cannot be fixed 
reliably without an intrinsic determination of the scale $\mu$. 

The present framework allows for a number of further invstigations. For instance, 
the question of  possible deconfinement phase transition in our approach 
should be answered from the effective action for the Polyakov loop, which gives a 
much stronger argument than the $O(4)$-symmetry violation in the propagators alone. 
This investigation is currently underway. In addition, the inclusion of fermions and 
a chemical potential is possible without further conceptional obstacles. To obtain 
realistic results, it may, however, be necessary to go beyond the Gaussian ansatz 
used in the present paper. This can be achieved e.g.~with the Dyson-Schwinger method 
used in the context of variational calculations within the Hamiltonian 
approach in Coulomb gauge~\cite{Campagnari:2010wc}.

%%%%%%%%%%%%%%%%%%%%%%%%%%%%%%%%%%%%%%%%%%%%%%%%%%%%%%%%%%%%%%%%%%%%%%%%%%%%%%%%%%%%%%%

\begin{acknowledgments}
The authors would like to thank A.~Sternbeck for providing the lattice data used in 
section \ref{sec:5}. This work was supported by DFG under contract Re-856/9-1. 
\end{acknowledgments}

%%%%%%%%%%%%%%%%%%%%%%%%%%%%%%%%%%%%%%%%%%%%%%%%%%%%%%%%%%%%%%%%%%%%%%%%%%%%%%%%%%%%%%%%

\appendix
\section{Explicit calculations}
\subsection{The zero-temperature gap equation}
\label{app:zero}
In ref.~\cite{Quandt:2013wna} the (unrenormalized) gap equation at zero temperature 
was derived in the form 
\begin{alignat}{3}
\bar{\omega}_0(\ka) &= k^2 + \chi_0(\ka) + M_0^2& \qquad & \qquad &
M_0^2 &= N g^2\,I_M^{(0)}
\nonumber \\[2mm]
\chi_0(\ka) &= N g^2\,I_\chi^{(0)}(\ka)  & \qquad & \qquad &
\eta_0(\ka)^{-1} &= 1 - N g^2\,I_\eta^{(0)}(\ka)\,,
\label{501}
\end{alignat}
where the index '0' on all quantities indicates that these are the zero-temperature 
profiles. The integrals in this equation are given by
\begin{alignat}{2}
I_M^{(0)}  &\equiv\,\,&\frac{9}{4}&\int \frac{d^4 q}{(2\pi)^4}\,\frac{1}{q^2 + M_0^2 + \chi_0(\qa)}
\nonumber \\[2mm]
I_\eta^{(0)}(\ka) &\equiv\,\, &&\int \frac{d^4 q}{(2\pi)^4}\, \Big[1 - (\unitfour{k}\cdot
\unitfour{q})^2\Big]\,\frac{\eta_0(k-q)}{(k-q)^2\,\bar{\omega}_0(q)}
\nonumber \\[2mm]
I_\chi^{(0)}(\ka) &\equiv\,\,&\frac{1}{3}& \int \frac{d^4 q}{(2\pi)^4}\,\Big[1 - 
(\unitfour{k}\cdot \unitfour{q})^2\Big]\frac{\eta_0(k-q)\,\eta_0(\qa)}{(k-q)^2}\,.
\label{502}
\end{alignat}

\subsection{Zero temperature mass functions}
\label{app:mass}
Here, we want to show the zero temperature limit
\begin{align}
\lim_{\beta\to\infty} M_\perp^2(\beta)  = M_0^2\,,\qquad\qquad
\lim_{\beta\to\infty} \Ml^2(\beta)  = 0
\label{app:100}
\end{align}
of the mass functions that appear in the gap equation (\ref{420}) and (\ref{430}) in the 
main text. We employ the auxiliary relation 
\begin{align}
\lim_{\beta\to\infty}\int_\beta \dd q\,f(q^2) \,\frac{\vek{q}^2}{q^2} = 
\frac{3}{4}\,\int\frac{d^4 q}{(2\pi)^4}\,f(q^2)\,,
\label{app:102}
\end{align}
which follows from $O(4)$-invariance in the limit $\beta\to\infty$.
% we prove first: From $O(4)$ invariance, we have
% \begin{align*}
% \lim_{\beta\to\infty} \int \dd q\,f(q^2)\,\frac{q_\mu q_\nu}{q^2} = 
% \int\frac{d^4 q}{(2\pi)^4}\,f(q^2)\,\frac{q_\mu q_\nu}{q^2} = \delta_{\mu\nu}\,K
% \end{align*}
% with some constant $K$ that can be computed by contracting indices,
% \begin{align*}
% 4 K = \int \frac{d^4 q}{(2\pi)^4}\,f(q^2)\,. 
% \end{align*}
% Contracting instead with the spatial projector $\mathcal{T}_{\mu\nu} = \text{diag}(0,1,1,1)$, 
% we find
% \begin{align*}
% \lim_{\beta\to\infty}\int_\beta \dd q\,f(q^2) \,\frac{\vek{q}^2}{q^2} =
% \mathcal{T}_{\mu\nu}\,\lim_{\beta\to\infty} \int \dd q\,f(q^2)\,\frac{q_\mu q_\nu}{q^2}
% = \mathcal{T}_{\mu\nu}\,\delta_{\mu\nu}\,K = 3 K = 
% \frac{3}{4}\,\int\frac{d^4 q}{(2\pi)^4}\,f(q^2)\,.
% \end{align*}
If we use this relation and take the zero-temperature limit eq.~(\ref{506}) into account, 
we find 
\begin{align*}
\lim_{\beta\to\infty} M_\perp^2(\beta) &= \frac{Ng^2}{2}\,\int\frac{d^4q}{(2\pi)^4}\,
\frac{A + B(\qa)}{\bar{\omega}_0(\qa)} = \frac{Ng^2}{2}\,\int\frac{d^4q}{(2\pi)^4}\,
\frac{\frac{14}{3} - \frac{2}{3}\,\frac{q_0^2}{q^2}}{\bar{\omega}_0(\qa)}
\\[2mm] 
&= Ng^2\,\int\frac{d^4q}{(2\pi)^4}\,\frac{1}{\bar{\omega}_0(\qa)}\,
\left[2 + \frac{1}{3}\, \frac{\vek{q}^2}{q^2}\right]
\\[2mm]
&\stackrel{(\ref{app:100})}{=} 
Ng^2\,\int\frac{d^4q}{(2\pi)^4}\,\frac{1}{\bar{\omega}_0(\qa)}\,\left[2 + \frac{1}{3}\cdot\frac{3}{4}\right]
\\[2mm]
&\stackrel{(\ref{502})}{=} M_0^2\,.
\end{align*}
The second equation (\ref{app:100}) follows in exactly the same manner.

\subsection{Loop integrals at finite temperature}
\label{app:loop}
The integrals in the main result eq.~(\ref{1000}) take the explicit form
\begin{align}
I_{\eta}(k_0,|\vek{k}|)   &=\int_\beta\dd q\,\frac{\eta(k-q)}{(k-q)^2 }\,
\frac{1 - (\unitfour{k}\cdot\unitfour{q})^2}{\sb(\qa)}
\nonumber\\[2mm]
L_{\eta}(k_0,|\vek{k}|)   &= \int_\beta\dd q \,\frac{\eta(k-q)}{(k-q)^2 }\,
\big(1 - (\unitthree{k}\cdot\unitthree{q})^2\big)\,
\Big [\wb(\qa)^{-1} - \sb(\qa)^{-1}\Big]
\nonumber\\[2mm]
I_\chi^\perp(k_0,|\vek{k}|) &=\frac{1}{2}\int_\beta \dd q\,\frac{\eta(k-q)\,
\eta (\qa)}{(k-q)^2}\,
\frac{\vek{q}^2}{q^2}\,\Big[ 1 - \big(\unitthree{k}\cdot\unitthree{q}\big)^2 \Big] 
\nonumber\\[2mm]
I_\chi^{\|}(k_0,|\vek{k}|)   &= \int_\beta \dd q\,\frac{\eta(k-q)\,
\eta(\qa)}{(k-q)^2}\,\left[\frac{q_0^2 + \mathbf{q}^2\,
\big(\unitthree{k}\cdot \unitthree{q}\big)^2}{q^2} - (\unitfour{k}\cdot\unitfour{q})^2\right] 
\nonumber\\[2mm]
M_{\perp}^2(\beta) &= \frac{1}{3}\,Ng^2\,\int_\beta \dd q\,\left[\frac{4}{\wb(\qa)}
+ \left(\frac{2 q_0^2 + 3 \vek{q}^2}{q_0^2 + \vek{q}^2}\right)\,
\frac{1}{\sb(\qa)}\right]
\nonumber\\[2mm]
% M_0^2 &= \frac{1}{2}\,Ng^2\,\int \frac{d^4 q} {(2\pi)^4}\,\left[\frac{A + B(\qa) }{\bar{\omega}_0(\qa)}\right]
% \nonumber\\[2mm]
\Ml^2(\beta)  &= \frac{1}{3}\,Ng^2\,\int_\beta \dd q
\left[ \frac{2}{\wb(\qa)} + 
\left( \frac{q_0^2 - 3 \vek{q}^2}{q_0^2 + \vek{q}^2}\right)\,
\frac{1}{\sb(\qa)}\right]\,.
\label{1001}
\end{align}
For simplicity, we have written $\eta(q)$ etc.~for the various form factors inside 
the loop integrals, although they depend, of course, on the two invariants $q_0$ and 
$|\vek{q}|$ separately. 

\subsection{Renormalization constants at finite temperature}
\label{app:constants}
As stated in the main text, the renormalized integral equation system 
eq.~(\ref{800x}) contains five unknown, temperature-dependent constants which 
are all determined by zero-temperature counter terms. In detail, these relations 
are
\begin{alignat}{2}
\eta(0,\mu_c)^{-1} &= \eta_0(\mu_c)^{-1} - \Big[ I_\eta(0,\mu_c) - I_\eta^{(0)}(\mu_c)
 - L_\eta(0,\mu_c) \Big] & & \,=\,\mu_0(\mu_c)^{-1} + \cdots
\nonumber \\[2mm]
\omega_\perp(0,\mu) &= Z\,\mu^2 + M_\perp^2(\beta) - M_0^2 + I_\chi^{(0)}(\mu) - I_\chi^\perp(0,\mu)
&  &\,=\, Z\,\mu^2 + \cdots 
\nonumber \\[2mm]
\omega_{\|}(0,\mu) &= Z\,\mu^2 + M_\perp^2(\beta) - M_0^2 + M_{\|}^2(\beta) + I_\chi^{(0)}(\mu) - 
I_\chi^{\|}(0,\mu) 
& & \,=\, Z\,\mu^2 + \cdots
\nonumber \\[2mm]
\omega_\perp(0,\mu_0) &= Z\,M_A^2 + M_\perp^2(\beta) - M_0^2 + I_\chi^{(0)}(\mu_0) - I_\chi^\perp(0,\mu_0)
&  &\,=\, Z\,M_A^2 + \cdots 
\nonumber \\[2mm]
\omega_{\|}(0,\mu_0) &= Z\,M_A^2 + M_\perp^2(\beta) - M_0^2 + M_{\|}^2(\beta) + I_\chi^{(0)}(\mu_0) - 
I_\chi^{\|}(0,\mu_0) 
& & \,=\, Z\,M_A^2 + \cdots\,,
\label{wurg}
\end{alignat}
where the dots in the last form are the finite-temperature corrections to the 
zero-temperature renormlization data. These corrections are very hard to treat numerically 
because the corresponding difference of loop integrals involves the finite-temperature 
and zero-temperature form factors, respectively. 

%%%%%%%%%%%%%%%%%%%%%%%%%%%%%%%%%%%%%%%%%%%%%%%%%%%%%%%%%%%%%%%%%%%%%%%%%%%%%%%%%%%%%%%%

\bibliographystyle{apsrev4-1}
\bibliography{varT}
\end{document}